# Sketched Nanoscale KTaO$_3$-Based Superconducting Quantum Interference Device

Muqing Yu[1], Nicholas Hougland[1], Qianheng Du[2], Junyi Yang[2], Sayanwita Biswas[1], Ranjani Ramachandran[1], Dengyu Yang[1,3], Anand Bhattacharya[2], David Pekker[1], Patrick Irvin[1], Jeremy Levy[1,*]

[1]*Department of Physics and Astronomy, University of Pittsburgh*
[2]*Materials Science Division, Argonne National Laboratory*
[3]*Department of Physics, Carnegie Mellon University*
* jlevy@pitt.edu

The discovery of two-dimensional superconductivity in LaAlO$_3$/KTaO$_3$ (111) and (110) interfaces has raised significant interest in this system. In this manuscript we report the first successful fabrication of a superconducting quantum interference device (DC-SQUID) in the KTO system. The key device elements, superconducting weak links, are created by conductive atomic force microscope (c-AFM) lithography which can reversibly control the conductivity at the LAO/KTO (110) interface with nanoscale resolution. The periodic modulation of the SQUID critical current, $I_c(B)$, with magnetic field corresponds well with our theoretical modeling, which reveals a large kinetic inductance of the superconducting two-dimensional electron gas (2DEG) in KTO. The kinetic inductance of the SQUID is tunable by electrical gating from the back, due to the large dielectric constant of KTO. The demonstration of weak links and SQUIDs in KTO broadens the scope for exploring the underlying physics of KTO superconductivity, including the role of spin-orbit-coupling, pairing symmetry, and inhomogeneity. It also promotes KTO as a versatile platform for a growing family of quantum devices, which could be applicable in the realm of quantum computing and information.
**Keywords:** nanoelectronics, superconducting weak link, SQUID, complex oxide, c-AFM lithography

## Introduction

The recent discovery of superconductivity at KTaO$_3$ (KTO) interfaces[1–4] reignites interest in very low electron density superconductors, also called superconducting semiconductors, and creates new opportunities for developing future quantum technologies. In comparison with the more extensively studied SrTiO$_3$ (STO)[5,6], KTO shows significantly stronger spin-orbit coupling (SOC) associated with the 5*d* Ta conduction band[7,8]. Unlike STO, KTO does not show any signature of superconductivity in its bulk form[9]. However, two-dimensional electron gas (2DEG) at KTO surface [10] or heterointerface with other materials can host superconductivity, with $T_c$ up to 2 K observed at KTO (111) interface [1], an order of magnitude larger than $T_c$ of STO. $T_c$ of the KTO 2DEG strongly depends on its crystallographic orientation, which may hold clues about its origin[1,2,11,12], and which may be useful in quantum device development. The combined properties of KTO 2DEGs make them natural candidates for hosting *p*-wave superconductivity[13] and Majorana zero modes[14–16] in KTO devices.

Superconducting weak links (WL) and Josephson junctions[17,18] in particular are the key nonlinear elements of superconducting quantum devices, from qubits to quantum-limited parametric amplifiers[19–21]. Moreover, the behavior of WLs can help pinpoint the underlying properties of the superconducting condensate[22,23]. However, a WL is yet to be created in KTO systems. Recently, it was demonstrated that conductive atomic force microscope (c-AFM) lithography can be used to pattern oxide 2DEG in LAO/KTO heterostructures[24,25] with nanoscale resolution "on demand", similar to previous work on LAO/STO heterostructures[26]. Based on this technique, a wide variety of mesoscopic devices such as single-electron transistors (SETs)[27], nanoscale photodetectors[28,29], and electron waveguides[30] were realized in LAO/STO.

In this paper, we report creation of nanoscale WLs (Dayem bridge type) at superconducting LAO/KTO (110) interface, using c-AFM lithography. By arranging two WLs in parallel, we are able to create superconducting quantum interference devices (SQUIDs) on KTO. The current-phase relationship of our WLs is analyzed by modeling the SQUID, which also serves as a sensitive magnetometer. The development of SQUID devices portends a wide range of superconducting devices based on the reconfigurable LAO/KTO interface.

**Results and Discussion**

We use c-AFM lithography to "sketch" mesoscopic SQUIDs at the LAO/KTO (110) heterointerface. Details of the c-AFM technique are given in the Methods section. We describe the behavior of four SQUIDs, which are labeled Device A, B, C and D. In the main text, we focus our discussion on Devices A and B. Data for Device C and D are presented in Fig.S13 and Fig.S14.

Fig.1 illustrates representative results from Device A. Fig.1(a) shows the layout of the device as sketched by c-AFM overlaying the AFM topographic image of the sample. The central part of the device (Fig.1(b)) consists of a square annular region measuring of 1.2 μm × 1.2 μm (outer area $A_o =$ 1.44 µm$^2$) and an inner hole measuring 0.4 µm × 0.4 µm (inner area $A_i = 0.16$ µm$^2$). The annulus is "cut" in half with the c-AFM using a negative tip voltage along the red lines, separating the top and bottom half of the annulus. A pair of Dayem bridges is subsequently sketched with positive voltage, as indicated by the lighter green features in Fig.1(b). The pair of bridges serves as two WLs for the SQUID. Each of the top two corners of the device is contacted by a single lead, while the bottom two corners are each contacted by a pair of leads. The step-by-step procedure used to create Device A is laid out in Fig.S4(a)-(e). Current-voltage ($I - V$) measurements are performed by sourcing current from the $I_+$ lead to the $I_-$ lead, while measuring the voltage across the $V_+$ and $V_-$ leads such that $V = V_+ - V_-$. In Fig.1(c), the $I - V$ characteristics and associated differential resistance $dV/dI$ of the SQUID are measured at $T = 50$ mK and magnetic field $B = 0$ T, from which a two-step transition can be seen. At small bias current, a low-resistance state is observed, with a residual resistance $dV/dI = 100$ Ω. As $I$ increases above a first critical current $I_{c,in} = 55$ nA (marked by green dashed lines, Fig.1(c)), $dV/dI$ gradually rises to approximately 2 kΩ. As $I$ continues to increase, a second critical current is identified $I_{c,out} = 80$ nA (marked by black dashed lines, Fig.1(c)) where $V$ shows a sudden jump, resulting in a sharp peak in $dV/dI$. We define $I_{c,in}$ to be the bias current where $dV/dI$ exceeds 550 Ω, which is half of the normal state resistance $R_n = 1.1$ kΩ (Fig.S10(a)), and define $I_{c,out}$ as the bias current where $dV/dI$ peaks. The regime between the two critical currents $I_{c,in} < I < I_{c,out}$ is characterized by a periodic dependence of the device resistance on the magnetic field (Fig.1(f)). In this regime, there must be phase slip fluctuations in the two Dayem bridges because we observe finite resistance. However, because we observe magnetoresistance oscillations, there must be coherence along a contour which crosses the two bridges and encircles the hole of the SQUID. This coherence implies that phase slips in the two bridges are correlated [31]. On the other hand, above $I_{c,out}$, the device resistance no longer demonstrates periodic dependence on magnetic field, and thus phase coherence must be disrupted by the fluctuating superconductivity. We remark that the linear extension of the $I - V$ curve in this regime does not cross zero, and therefore we do not believe this regime to be a fully normal state (Fig.S6(a)). We also note that a small hysteresis in $I - V$ curve is observed near $I_{c,out}$ (Fig.S6(a)). The magnitude of hysteresis is negligible (< 1%) compared to $I_{c,out}$, so we choose the backward sweep (red curve, Fig.S6(a)) to represent the $I - V$ characteristic of Device A.

Fig.1(d,e) shows waterfall plots of $V$ vs $I$ and $dV/dI$ vs $I$ curves as a function of out-of-plane magnetic field $B$ ranging from ~ 0 Oe to ~ 55 Oe. Fig.1(f) shows the corresponding intensity plot of $dV/dI$, where $I_{c,in}$ and $I_{c,out}$ are marked by white and black dashed lines. $I_{c,out}$ monotonically decreases from 80 nA to 70 nA as $B$ increases from 0 to 200 Oe. We ascribe this decrease to the suppression of superconductivity by magnetic field. Meanwhile, the inner critical current $I_{c,in}$ oscillates between 45 nA

and 64 nA as $B$ changes, with a period $\Delta B = 27.4$ Oe. For a SQUID, $\Delta B = \Phi_0/A$, where $\Phi_0$ is the flux quantum and $A$ is the geometrical area enclosing the magnetic field. For Device A, we get an effective area $A_{\text{eff}} \sim 0.75$ μm² such that $A_i < A_{\text{eff}} < A_o$. This discrepancy results from the mesoscopic dimension of our superconducting leads. As the superconducting leads are significantly smaller than the Pearl length, the magnetic field penetrates the leads almost without attenuation. Consequently, the effective area of the device, as determined by the period of critical current oscillations with the magnetic field, is larger than the inner hole area $A_i$. We find the effective area by describing the screening currents (see Theory Modeling section and Refs.[31,32]). As the magnitude of $B$ increases past ~120 Oe, $I_{c,\text{in}}$ oscillations show discontinuities (Fig.1(f)), which we associate with the formation of vortices within the 2D superconducting regions[31]. Because we are mostly interested in the oscillating $I_{c,\text{in}}$ rather than $I_{c,\text{out}}$, in the following discussion we define $I_c$ to be $I_{c,\text{in}}$ unless specified.

The LAO/KTO SQUIDs are characterized by a high kinetic inductance, which dominates their behavior as will be described below. In Fig.2, we plot the positive and negative critical currents, $I_{c\pm}(B)$, of Device A as functions of magnetic field. We note that although $I_{c+}(0) = |I_{c-}(0)|$, the maxima of $I_{c+}$ and $|I_{c-}|$ do not occur at $B = 0$ T, and there is a phase shift between $I_{c+}$ and $I_{c-}$ versus magnetic field. We model the behavior of the SQUIDs using the methods described below in Theory Modeling. From this model, we are able to derive the effective length $l_{l(r)}$, critical current $I_{c,l(r)}$, critical phase $\varphi_{c,l(r)}$ and kinetic inductance $L_{k,l(r)}$ of the left (right) Dayem bridge, as well as the minimum superfluid density $n_s$ of the device. For Device A, we get $I_{c,l} = 26.4$ nA, $I_{c,r} = 34.0$ nA, $\varphi_{c,l} = 3.50$ rad and $\varphi_{c,r} = 4.81$ rad (green lines, Fig.2). This ~22% mismatch in critical current and ~25% mismatch in critical phase between the left and right bridges account for the experimentally observed phase shift between $I_{c+}(B)$ and $I_{c-}(B)$. We also get the effective length $l$ in units of Ginzburg-Landau coherence length $\xi$: $l_l = 6\xi$ and $l_r = 8.9\xi$. A separate Hall bar device suggests $\xi = 38$ nm (Fig.S3(e)), which gives estimated lengths $l_l = 230$ nm and $l_r = 340$ nm of the bridges. We note, this estimated coherence length may not directly apply to the case of the Dayem bridges as they are much narrower than the Hall bar. We also note that the overlapping region between the written nanowires (light green paths, Fig.1(b)) and the annulus (dark green, Fig.1(b)) can have higher carrier density compared to the rest of the 2D annulus. Also, the 2D annulus itself can host inhomogeneous superconductivity because some regions are written twice by c-AFM lithography (Fig.S4(b)). This inhomogeneity may result in the derived effective length being longer than the physical length of the bridge. The kinetic inductance of each bridge is $L_{k,l} = 30.7$ nH, $L_{k,r} = 33.6$ nH. $L_k$ of a superconducting nanobridge typically scales with its length $l$. Using the bridge lengths derived from the Hall bar measurement, we can establish a lower bound on the kinetic inductance per unit length of 100 nH/μm, which is ~2 orders of magnitude greater than $L_k$ of nanowires (bridges) made out of Nb and NbN (Ref.[33]), MoGe (Ref.[34]), and granular Al (Ref.[35]).

Because the KTO has a low-temperature dielectric constant $\varepsilon \sim 5000$ [36], and the carrier density of the KTO 2DEG is relatively low, i.e. $n_{2D} = 5.4 \times 10^{13}$ cm$^{-2}$ (Fig.S3(c)), our KTO-based SQUIDs are sensitive to electric gating. Fig.3(a) shows an intensity plot of $dV/dI$ against bias current and backgate voltage $V_{bg}$, where $I_c$ (white dashed lines) increases monotonically as $V_{bg}$ increases. When $V_{bg}$ increases past $\approx +20$ V, $I_c$ begins to saturate. Similar dependence of critical current on $V_{bg}$ has previously been reported in Ref.[24]. Fig.3(b)(c) show the intensity plot of $dV/dI$ versus $I$ and $B$ measured at $V_{bg} = -80$ V and $V_{bg} = +70$ V, respectively. In both cases, the $I_c$ oscillation period $\Delta B$ remains unchanged at 27.4 Oe. From our theory modeling, we can extract the bridge critical currents under $V_{bg} = -80$ V and $V_{bg} = +70$ V (Fig.3(d)), which are listed in Table 2 for comparison with $V_{bg} = 0$ V. At $V_{bg} = +70$ V, $I_{c,l} = 26.6$ nA and $I_{c,r} = 34.1$ nA, showing a slight increase compared to $V_{bg} = 0$ V. The effect of negative $V_{bg}$ is more pronounced: a $-80$ V backgate voltage reduces $I_c$ to $I_{c,l} = 18.2$ nA and $I_{c,r} = 25.5$ nA. In this case, the bridge kinetic inductances are $L_{k,l} = 44.6$ nH and $L_{k,r} = 43.7$ nH, a >40% increase when compared to $V_{bg} = 0$ V. This increase in kinetic inductance is consistent with decreasing superfluid density under negative backgate voltage.

To operate the SQUID as a flux-to-voltage transformer, its bias current $I$ is fixed while its voltage is modulated by magnetic flux. Fig.4(a) shows the voltage modulations, with bias current ranging from $I = 30$ nA to $I = 85$ nA, measured at $T = 50$ mK. We can extract the maximum of the flux-to-voltage transfer coefficient $|dV/d\Phi|$ for each bias current $I$ (Fig.4(b)). Max $|dV/d\Phi|$ shows non-monotonic dependence on $I$. The optimal bias current is $I = 66$ nA, where the transfer coefficient reaches up to 65 μV/$\Phi_0$. Fig.S7 shows intensity plots of $dV/dI$ at elevated temperatures. The magnitude of $I_c$ decreases as temperature increases, and its periodic dependence on magnetic field is completely smeared out when temperature increases past 0.5 K. There is a slight decrease of $I_c$ in Fig.S7(a) when compared to Fig.1(f), which is due to the hysteretic effect of $V_{bg}$ sweep during the backgate-dependent experiments described in the previous paragraph. Fig.4(c) shows the transfer function $|dV/d\Phi|$ vs $I$ over a range of temperatures, and Fig.4(d) plots the optimal $|dV/d\Phi|$ at each temperature. The optimal $|dV/d\Phi|$ decreases monotonically with increasing temperature, due to the suppression of superconductivity. The flux noise spectral density $S_\Phi$ is related to the voltage noise spectral density $S_V$ by $S_\Phi = S_V/|dV/d\Phi|$. At $T = 50$ mK, the intrinsic noise of our device mainly consists of thermal noise, which has a spectral density $S_V = \sqrt{4k_B T R} = 65$ pV/$\sqrt{\text{Hz}}$. This gives an expected noise sensitivity $S_\Phi \sim 1.0$ μ$\Phi_0/\sqrt{\text{Hz}}$, comparable to the sensitivity of conventional DC-SQUIDs ($S_\Phi = 1 \sim 10$ μ$\Phi_0/\sqrt{\text{Hz}}$, Ref.[37]). However, during the measurement, our voltage preamplifier gives a larger voltage noise of $S_V = 7$ nV/$\sqrt{\text{Hz}}$, which translates to $S_\Phi \sim 100$ μ$\Phi_0/\sqrt{Hz}$. Detailed noise measurements are needed to confirm the values of $S_V$ and $S_\Phi$. We note that in this device we observe unphysical periodic $V$ oscillations with $B$ field even at zero bias current, which is discussed in detail in supplementary note 1 and Fig.S17.

Device B (Fig.5(a)) is a DC-SQUID consisting of two Dayem bridges connecting the top and bottom halves of a 2D annulus. This device is similar to Device A, but has an area which is reduced by ~30%. It has a square annular area $A_o = 1.04$ μm$^2$ and an inner insulating area of $A_i = 0.12$ μm$^2$. The detailed procedure of creating Device B is laid out in Fig.S4(f)-(j). From the $I - V$ measurements below $T_c$, we again observe a two-step transition ($I_{c,in}$ and $I_{c,out}$, Fig.5(b)) as well as the periodic modulation of $I - V$ characteristics by magnetic field (Fig.5(c,d)). An intensity plot of d$V$/d$I$ vs $I$ and $B$ is shown in Fig.5(e), where we can extract the $I_c$ oscillation period to be $\Delta B = 32.6$ Oe, corresponding to an effective area $A_{eff} = 0.63$ μm$^2$. For Device B, we find a ~20% increase in $\Delta B$ compared to Device A, which is expected for a SQUID with smaller dimensions. Like Device A, the effective area of Device B is larger than its inner insulating area, again due to Meisner current within the 2D annulus. From theoretical fit, we get $I_{c,l} = 40.8$ nA, $I_{c,r} = 46.5$ nA, $\varphi_{c,l} = 5.68$ rad and $\varphi_{c,r} = 5.83$ rad (Fig.5(f)). The two bridges of Device B have similar critical currents (~13% mismatch) and similar critical phases (~2% mismatch), resulting in a "typical" SQUID behavior where the maxima of $I_{c+}$ and $|I_{c-}|$ are both reached at $B = 0$ T. The kinetic inductance of the two arms is $L_{k,l} = 32.3$ nH and $L_{k,r} = 29.1$ nH, which is in the same order as $L_k$ of Device A (at $V_{bg} = 0$ V). Each bridge we fabricated has slightly different critical current and kinetic inductance, which could result from a change in c-AFM lithography parameters (Fig.S4). A detailed understanding of how c-AFM lithographic parameters impact the current-phase relationship of nanowires at low temperature is worthy of investigation but is yet to be established. We also investigated Device B when used as a flux-to-voltage transformer. At $T = 50$ mK, voltage modulation at certain values of bias current is plotted in Fig.S8(a), and corresponding max $|dV/d\Phi|$ is extracted and plotted in Fig.S8(b). With $I = 90$ nA, Device B has an optimal $|dV/d\Phi| = 190$ μV/$\Phi_0$, three times that of Device A. During the measurement of Device B, we switched current source and drain to different leads, and corresponding $I_c(B)$ relationships are shown in Fig.S9. For these different measurement configurations, $I_{c,l(r)}$ of each bridge are extracted and listed in Table 3. We note that $I_{c,l(r)}$ differ depending on the choice of leads. This lead dependence could result from an electrical gating effect based on the choice of current leads, or from unintended inhomogeneities in superfluid density within the 2D annulus (Fig.S4(f)-(j)).

In some of our SQUIDs (Device A, B and D), as well as single-Dayem-bridge devices (Device E & F), we observe a residual resistance down to the lowest temperature we can reach. For example, this residual resistance is reflected in the slightly slanted $I - V$ curve of Device A and B near zero bias at $B = 0$ T (Fig.1(c), Fig.5(b)). We observe that the residual resistance becomes suppressed at finite magnetic fields and slightly elevated temperatures, a phenomenon which we will discuss in the next paragraph. Here, we comment on the origin of the residual resistance, and its relation to the observed critical current oscillations with magnetic field. According to LAMH theory, thin superconducting wires always carry a finite resistance even below $T_c$, as thermal fluctuations drive some point in the wire to normal and thus cause phase slippage [38,39]. Although our PPMS instrument comes with a RF filter (see Methods section), some RF noise may still be reaching the device being measured and driving phase slips in it. In thin wires there are also possibilities for quantum phase slips [40], which can cause resistance even at $T = 0$. We suspect that a combination of these phase-slip mechanisms happens in our measurement of KTO Dayem bridges. However, even in the presence of phase slips, SQUIDs can still be phase coherent [31,32]. Indeed, from our observation of quantum interference in our KTO SQUIDs we can infer that: (1) There is long range phase coherence in the top and bottom half of our annulus; (2) Supercurrent along our two bridges is phase-dependent. Further, the quantum interference pattern we observe relies on the existence of the superconducting state, as it disappears at $T > 600$ mK when KTO transitions to normal metal (Fig.S7). Therefore, we rule out the single-electron Aharonov-Bohm effect [41]. Thus, we believe it is justified to call Device A and B superconducting quantum interference devices (SQUIDs).

Here we discuss the temperature and field dependence of the resistance in our Dayem bridges. As shown in Fig.S10(a), Device A goes through a broad superconducting transition at $0.5$ K $< T < 1$ K. As we continue to lower the temperature below 0.4 K, we see an unexpected feature: the resistance increases and then remains finite down to 50 mK. SQUIDs B and D as well as the single Dayem bridges (Device E and F) show similar non-monotonic resistance-temperature relationships (Fig.S11(a), S14(d), S15(c) and S16(b)). In these devices, the resistance plateauing at a finite value when cooled close to 0 temperature is different from the standard Josephson junctions, which becomes either zero or infinite resistance at zero temperature [42,43]. We note SQUID C shows monotonic decrease of resistance with temperature (Fig.S13(b)), and its resistance is immeasurably small at $T = 50$ mK. Next, we investigate the resistance as a function of magnetic field at $T = 50$ mK. Fig.S10(b) shows the zero-bias resistance of Device A versus magnetic field, where a resistance peak spanning from $-0.07$ T to $+0.07$ T is observed. In single Dayem bridges (Device E and F), we observed similar residual resistance which can be suppressed by slightly elevated magnetic field (Fig.S15(e) and Fig.S16(e)). SQUID B also has a residual resistance peak near $B = 0$, although it can be suppressed by a small $B$ field $|B| = 50$ Oe (Fig.S11(d)). This resistance peak exhibits hysteresis with respect to $B$ field. The critical current vs $B$ oscillations of SQUID A and B also show hysteresis under $B$ field, which is discussed more in supplementary note 2 and Fig.S11 and Fig.S12. SQUID C and D, different from other devices, do not show any resistance peak near $B = 0$ T (Fig.S13(e), FigS14(g)). We note that SQUID D has a zero-bias resistance peak in all field range (Fig.S14(b)). As non-monotonic resistance-temperature curves and zero-field resistance peaks can occur in both SQUID and single Dayem bridge devices we ascribe these effects to transport phenomena in individual Dayem bridges rather than the interference between the two bridges in SQUIDs. Neither LAMH theory nor quantum phase slips can explain these behaviors. One possible reason for the zero-field resistance peak is magnetic impurities within the KTO substrate, which induce scattering at $B = 0$ T and would be suppressed when the magnitude of $B$ increases [44–46]. Sscattering from magnetic impurities can be more prominent in ultranarrow wires [46], which may cause the zero-field resistance peak to be obvious in our KTO Dayem bridges as compared to the KTO 2DEG in general. There are also reports of ferromagnetism in KTO [47,48], which could explain the origin of magnetic impurities. Magnetism in KTO can also result in the observed $B$ field hysteresis in SQUIDs A & B. Another possible explanation would be a phase transition, perhaps from an s-wave to a p-wave superconducting state in KTO (Ref.[13]). Further investigations are needed to identify the source of this resistance anomaly. Concerning the zero-bias resistance peak in SQUID D, we suspect self-formed charging islands [49].

**Theory Modeling**

Here we describe the theoretical model for the DC-SQUID which we use to understand the magnetic field period and to fit critical current oscillations and extract device parameters. We model the two half-annuli of the SQUID as uniform 2D superconductors. As the annuli are much smaller than the Pearl length, external magnetic fields penetrate them almost without attenuation. However, such fields generate screening currents and associated phase gradients. We model the two Dayem bridges as short superconducting nanowires. Oscillations of the device critical current with magnetic field result from phase quantization around a contour that encircles the inner square of the annulus. We model the free energy in the KTO annulus due to the screening supercurrents as

$$F = \int \frac{1}{2m^*} |\psi|^2 \left| \hbar \nabla \varphi - \frac{e^*}{c} A \right|^2 dx^2 \qquad (1)$$

where $A$ is the magnetic vector potential, $\varphi$ is the phase of the superconducting order parameter, $e^* = 2e$, $m^* = 2m$, and $|\psi|^2 = n_s$, the superconducting electron density. We supplement this equation by Neumann boundary conditions which prevent supercurrents from leaving the annuli, except at the points where the leads and Dayem bridges are attached, where the boundary conditions specify the currents. We relate the current across the Dayem bridges to the phase differences that appear across the half-annuli using the current-phase relationship for short superconducting nanowires[18,31,50]. This current-phase relationship depends on the length of the Dayem bridge, exhibiting Josephson junction-like behavior for short lengths and linear behavior for longer bridges, as observed in Ref.[51]. See Methods section for details of our theory modeling.

| Periodicity of $I_c$ vs B | Device A | Device B |
|---:|:---:|:---:|
| Experiment | 2.74 mT | 3.26 mT |
| Flux through hole | 12.9 mT | 17.9 mT |
| Our theory | 3.59 mT | 4.88 mT |

Table 1. The period of $I_c$ vs B oscillation of SQUIDs A and B (experiment), along with the naïve prediction from flux through hole, and the prediction from our theory.

We find the magnetic field period by minimizing the free energy given by Eq. 1 and applying phase quantization around the annulus. Table 1 compares the experimentally determined periodicity of critical current oscillations versus magnetic field to the estimated periodicity from our model of the superconducting half-annuli, as well as the naïve estimate from the area of the hole in the annulus. The naïve estimate exceeds the observed periodicity by a factor of roughly 4. Our model, which accounts for the penetration of the SQUID by external magnetic field and resulting phase variation across the annular regions of the device, determines this periodicity to within approximately 30% for Device A and 50% for Device B. We ascribe these discrepancies to the nonuniform superconductivity and imperfect knowledge of the device geometry resulting from the method in which the devices were fabricated. For example, some regions within the annulus of the SQUID are sketched twice, while other regions are sketched and then erased (Fig.S4), which may result in spatial variations of the device's superconducting properties.

For each device, we use our model of the half-annuli coupled to our model of the Dayem bridges to fit $I_{c,\text{in}}(B)$ by setting the critical currents and lengths of the bridges as well as the superfluid density of the half-annuli. As we do not know how much magnetic flux is trapped, we also fit an overall magnetic field offset. This model replicates the size of $I_{c+}$ and $I_{c-}$ oscillations with magnetic field, as well as their mutual phase offset. In particular, we are able to adjust the bridge critical current and length in order to tune these characteristics. Additionally, we find that our model provides a range of values of superfluid density $n_s$ which give physically similar results. That is, we are able to reduce the superfluid density while maintaining nearly identical results down to a minimum density, at which point the model no longer finds agreement with the remaining fitting parameters. We are also able to extract the critical phase and kinetic inductance from the current-phase relationship which is fitted for each Dayem bridge (see Theory Methods and Fig.S18). In Table 2, we provide the length, critical current, critical phase, and kinetic inductance of each bridge as determined by our model, as well as the minimum possible superfluid density for Device A with different applied backgate voltages. Similarly, in Table 3 we provide these parameters for Device B with currents applied at different leads. We provide a fit in each case since we find that the critical current of the device, and consequently of its Dayem bridges, varies depending on the orientation of the leads with applied current (Fig.S9).

As our Dayem bridges are equivalent to superconducting nanowires with width $w$, we can extract their $n_s$ from $I_c$ (see Theory Methods for details). In this calculation of $n_s$, we assume $w = 20$ nm, which is the resolution of c-AFM writing [24]. This value of $n_s$ is largely independent of the theory model described above, depending primarily on the measured values of $I_c$, $B_{c2}$, and $w$. This derivation also relies on the value of $l/\varphi_c$ which we extract from the theory model, but this parameter does not vary significantly since the bridge lengths are generally around 100 nm and the critical phases are largely between $\pi$ and $2\pi$. We find that $n_s$ of the bridges tends to be slightly lower than the minimum $n_s$ of the annulus given by the theory model, which is on the order of $10^{12}$ cm$^{-2}$ (Tables 2 and 3). This is consistent with weak links being written by lower c-AFM tip voltage compared to the annulus (Fig.S4). We also note that $n_s \sim 10^{12}$ cm$^{-2}$ is much lower than the total carrier density ($n_{2D} = 5.4 \times 10^{13}$ cm$^{-2}$, Fig.S3(c)) as obtained from Hall measurement. To our knowledge there is no other report on the superfluid density or kinetic inductance on the KTO (110) 2DEG. Ref.[4] reports a superfluid density of $n_s = 1.8 \times 10^{12}$ cm$^{-2}$ (which corresponds to kinetic inductance per square area of 1.1 nH/sq) and a total carrier density $n_{2D} = 7.5 \times 10^{13}$ cm$^{-2}$ in KTO (111) 2DEG. These values are on the same order as what we observe in KTO (110), but we note these are two different KTO systems with different $T_c$.

| **Device A** | $V_{bg} = -80$ V | $V_{bg} = 0$ V | $V_{bg} = +70$ V |
|---|---|---|---|
| $l_l/\xi_{GL}$ | 6.0 | 6.0 | 6.0 |
| $l_r/\xi_{GL}$ | 8.7 | 8.9 | 8.9 |
| $I_{c,l}$ (nA) | 18.2 | 26.4 | 26.6 |
| $I_{c,r}$ (nA) | 25.5 | 34.0 | 34.1 |
| $\varphi_{c,l}$ (rad) | 3.50 | 3.50 | 3.39 |
| $\varphi_{c,r}$ (rad) | 4.68 | 4.81 | 4.83 |
| $L_{k,l}$ (nH) | 44.6 | 30.7 | 30.6 |
| $L_{k,r}$ (nH) | 43.7 | 33.6 | 33.5 |
| Minimum annulus $n_s$ ($10^{11}$ cm$^{-2}$) | 8.6 | 14 | 14 |
| $n_{s,l}$ ($10^{11}$ cm$^{-2}$) | 3.18 | 4.61 | 4.82 |
| $n_{s,r}$ ($10^{11}$ cm$^{-2}$) | 4.82 | 6.44 | 6.41 |

Table 2. Length $l_{l(r)}$, critical current $I_{c,l(r)}$, critical phase $\varphi_{c,l(r)}$, and kinetic inductance $L_{k,l(r)}$, of each bridge in SQUID A, determined by theoretical fit. Minimum $n_s$ in the annulus given by theory model, and $n_s$ in the bridges calculated from $I_c$, are also listed. Different columns represent different backgate voltages applied.

| **Device B** | I+: left, I-: left Fig.S9(a) | I+: right, I-: right Fig.S9(b) | I+: right, I-: left Fig.S9(c) | I+: left, I-: right Fig.S9(d) |
|---|---|---|---|---|
| $l_l/\xi_{GL}$ | 10.2 | 5.1 | 12.0 | 10.6 |
| $l_r/\xi_{GL}$ | 10.5 | 8.8 | 8.7 | 4.9 |
| $I_{c,l}$ (nA) | 40.8 | 19.3 | 22.0 | 25.3 |
| $I_{c,r}$ (nA) | 46.5 | 24.3 | 27.8 | 35.1 |
| $\varphi_{c,l}$ (rad) | 5.68 | 3.13 | 6.52 | 5.76 |
| $\varphi_{c,r}$ (rad) | 5.83 | 4.73 | 4.72 | 3.04 |
| $L_{k,l}$ (nH) | 32.3 | 36.3 | 49.5 | 54.1 |
| $L_{k,r}$ (nH) | 29.1 | 46.5 | 40.2 | 19.4 |
| Minimum annulus $n_s$ ($10^{11}$ cm$^{-2}$) | 22 | 3.2 | 2.5 | 6.7 |
| $n_{s,l}$ ($10^{11}$ cm$^{-2}$) | 7.49 | 3.18 | 4.14 | 4.56 |
| $n_{s,r}$ ($10^{11}$ cm$^{-2}$) | 8.57 | 4.62 | 5.24 | 5.73 |

Table 3. Length $l_{l(r)}$, critical current $I_{c,l(r)}$, critical phase $\varphi_{c,l(r)}$, and kinetic inductance $L_{k,l(r)}$, of each bridge in SQUID B, determined by theoretical fit. Minimum $n_s$ in the annulus given by theory model, and $n_s$ in the bridges calculated from $I_c$, are also listed. Different columns correspond with different measurement configurations, which are depicted in Fig.S9(a)(b)(c)(d).

**Outlook**

We have successfully created superconducting WLs and DC-SQUIDs at LAO/KTO interface. The $I_c$ vs $B$ relationship of the SQUIDs reflects the high kinetic inductance of these WLs. Such large kinetic inductance makes KTO an intriguing platform for superconducting nanoinductors and RF superconducting circuits in general. It may also allow us to develop KTO fluxonium qubit which is insensitive to charge noise[52,53]. Moreover, the electrostatic tunability of kinetic inductance makes KTO-based electronics more adaptive and versatile. Because of the high resolution of c-AFM lithography, SQUIDs created with this technique can reach sub-micrometer scale while maintaining good sensitivity. The reconfigurability of c-AFM lithography promotes KTO SQUIDs as magnetic field sensors that can be programmed on-demand and integrated with magnetic nanostructures and 2D materials. The combination of superconductivity, spin-orbit coupling and nanoscale dimension in KTO nanodevices make it an intriguing candidate for exploring topological superconductivity and related devices.

## Methods

### Growth and Lithographic Patterning of LAO/KTO

A 4 nm amorphous LAO film is deposited onto KTO (110) substrate by pulsed laser deposition. LAO is removed everywhere except in a protected octagonal region via reactive ion etching (Fig.S1(a,b)). The middle 50 µm x 50 µm region of this "octagon" serves as the "canvas" on which superconducting structures will be created by c-AFM lithography.

### Conductive Atomic Force Microscope Lithography

Aluminum wirebonds are used to contact the LAO/KTO interface (Fig.S1(c)). The 4 nm LAO/KTO interface is found to be intrinsically conducting with a sheet conductance $G_{\text{sheet}} \sim 10 - 100$ uS/sq at room temperature. The conductive atomic force microscope lithography was performed using an Asylum Research MFP-3D AFM. The AFM chamber is maintained at ambient condition with a typical humidity around 40% and temperature of 27 °C. The DC writing voltage $V_{\text{tip}}$ was applied to the c-AFM tip (highly doped silicon) through a 1 GΩ tip resistor. The LAO/KTO interface can be rendered insulating and then switched back to conducting by c-AFM lithography (Fig.S2(a,b)). Prior to the creation of each SQUID, the central part of the canvas ($\sim 30\mu m \times 30\mu m$) is always erased into an insulating state (Fig.S2(c)). Fig.S4 describes the detailed procedures and c-AFM parameters for patterning SQUIDs. $V_{\text{tip}}$ in the range +30 V to +40 V were used to write the two-dimensional conducting regions, while $V_{\text{tip}}$ from -10 V to -12 V were used to erase (cut) the conducting regions back to insulating. The weak links are created with $V_{\text{tip}}$ between +8 V and +9 V. During c-AFM lithography, the four-terminal conductance of our devices is monitored in real-time (Fig.S5(a,b)). In ambient conditions, the lifetime of a sketched SQUID is typically around 10 h (Fig.S5(c)). At cryogenic temperatures, these devices are stable and have infinitely long lifetime, as reported by Ref.[24].

### Transport Measurements

Low-temperature transport measurements were carried out in a Quantum Design PPMS with a dilution refrigerator (DR) unit. The DR sample measurement cable comes with a RF filter (a ferrite-based RF choke) to attenuate RF signals outside of the DR probe. Within the DR probe, electrical connection to the sample is made by twisted pairs of manganin wires, which can further reduce the RF interference. However, we acknowledge there may still be unfiltered RF noises in our system. Source voltages were applied by a 24-bit digital/analog converter National Instruments PXI-4461, which can also simultaneously perform 24-bit analog/digital conversion. Current biasing was achieved by shunting the device with a 300 kΩ in-series resistor. The drain current and the voltages were measured after amplification by a Krohn-Hite Model 7008 multi-channel pre-amplifier. When taking one single I-V curve, the bias current starts from 0, then ramps up to positive maximum, then ramps down to negative minimum and finally back to 0, which takes ~20 s in total. At a certain magnetic field, 10~20 I-V curves are repeatedly taken and then averaged. A TD250 voltage amplifier is used for backgating purpose. Lock-in measurements are also applied to characterize the zero-bias resistance of our devices, especially during resistance versus temperature measurements. During lock-in measurements, an ac current $I = I_0 \sin(2\pi f t)$ passes through our devices, with $I_0 \approx 10$ nA and $f = 13$ Hz. In-phase component of the four-terminal voltage $X(V_{4T})$ is read out, so the four-terminal resistance is $R_{4T} = X(V_{4T})/I_0$.

**Theory Methods**

Solving for the current density in the annulus entails minimizing the energy and therefore solving

$$\hbar \nabla^2 \varphi = \frac{e^*}{c} \nabla \cdot A.$$

We use the Landau gauge $A = By\hat{e}_x$ with magnetic field $B = \hat{B}_z$ which is perpendicular to the plane of the device, and therefore $\nabla \cdot A = 0$. We supplement the Laplace equation with the Neumann boundary condition that $\left(\hbar \nabla \varphi - \frac{e^*}{c} A\right) \cdot \hat{n} = j$. The boundary normal current $j$ is selected to be 0 such that no current penetrates the boundaries of the device, except at the Dayem bridges and the current source leads where we specify $j$.

We discretize the resulting Laplace equation and solve it by successive over-relaxation on one half-annulus of the device. In order to derive a current-phase relation for the device, we consider the phases at the two Dayem bridges, $\varphi_1$ and $\varphi_2$, and the phase at the lead attachment point, $\theta$. Solving the Laplace equation relates these phases and the magnetic field to the currents $j_1$ and $j_2$ at the bridges. In particular,

$$j_1 = M_{11}(\theta - \varphi_1) + M_{12}(\theta - \varphi_2) + K_1 B$$
$$j_2 = M_{21}(\theta - \varphi_1) + M_{22}(\theta - \varphi_2) + K_2 B$$

where we fit the constants $M_{11}, M_{12}, M_{21}, M_{22}, K_1, K_2$ to solutions of the Laplace equation.

We incorporate the Dayem bridges using their current-phase relation $J_i(\Delta\varphi_i)$, where $\Delta\varphi_i$ is the phase difference across the $i$-th bridge. To obtain relation $J_i(\Delta\varphi_i)$, we start with the Ginzburg-Landau free energy:

$$F = \int_{-b/2}^{b/2} \alpha|\psi|^2 + \frac{\beta}{2}|\psi|^4 + \frac{\hbar^2}{2m}|\nabla\psi|^2 dx$$

subject to the following conditions:

$$|\psi(x = \pm b/2)|^2 = \frac{\alpha}{\beta},$$

$$\psi^*\nabla\psi + \psi\nabla\psi^* = J.$$

Taking $\psi = fe^{-i\varphi}$ and following the derivation in Ref.[31], we arrive at an expression for the phase across a wire carrying a current $J$. In particular,

$$\theta = \int_{-b/2}^{b/2} \frac{J}{f^2(x)} dx$$

$J(\Delta\varphi)$ calculated for bridges with different lengths $l$ (sampled from $l_l$ and $l_r$ in Table 2 & 3) are plotted in Fig.S18. When $l$ is greater than the Ginzburgh-Landau coherence length $\xi_{GL}$, $J(\Delta\varphi)$ deviates from the sinusoidal relation of typical Josephson junctions, and its critical phase $\varphi_c$ increases with $l$ (Fig.S18). For each bridge in the SQUID, we set $j_i = J_i(\Delta\varphi_i)$ to find the current-phase relationship of the SQUID. Using the derived relationships, we can relate the current into the device, $j_1 + j_2$, to the phase difference across the whole device, which we extract from the phase across both the bridges and the half-annuli. We then determine the critical current at a particular magnetic field by finding the maximum current supported at any phase difference across the entire SQUID.

The critical current of the bridges can be used to extract their superfluid density $n_s$, as follows

$$\frac{I_c}{w} = \frac{e^*\hbar}{m^*} n_s \frac{\varphi_c}{l}$$

where $\varphi_c$ is the critical phase of the bridge while $l$ and $w$ are its length and width, respectively. Our model provides a fit for $l/\xi_{GL}$, and so we can use this combined with the fact that $\xi_{GL} = \sqrt{\frac{\Phi_0}{2\pi B_{c2}}}$ to find $l$, where the upper critical field $B_{c2} = 0.23\ T$ (Fig.S3e). Combining the experimental parameters $I_c$ and $B_{c2}$ with the theory fit parameters $l/\xi_{GL}$ and $\varphi_c$, we compute $n_s$ for each bridge.

**Supplemental Material**

The following files are available free of charge:

Supplementary Information (PDF, 13 pages, including Supplementary Notes 1 & 2 and Supplementary Figures S1-S18)

**Acknowledgements**

**Funding:** This work was supported by NSF PHY-1913034 (J. L. and D.P.) and NSF DMR-2225888 (J.L.). All work at Argonne was supported by the US Department of Energy, Office of Science, Basic Energy Sciences, Materials Sciences and Engineering Division. The use of facilities at the Center for Nanoscale Materials and the Advanced Photon Source, both Office of Science user facilities, was supported by the US Department of Energy, Basic Energy Sciences under Contract No. DE-AC02-06CH11357. **Author's contributions:** J. L. supervised the experiments. Q. D., J. Y. and A. B. performed the KTO sample growth. M. Y. and D. Y. patterned the SQUIDs. M. Y., S.B., R. R. and P. I. conducted transport measurements at cryogenic temperature. M. Y. and N. H. analyzed the data. D. P. supervised theoretical modeling that was performed by N. H. All authors discussed the results and commented on the manuscript. **Completing interests:** The authors declare no competing interests. **Data Availability:** All data needed to evaluate the conclusions in the paper are present in the paper and/or the Supplementary Materials. Data necessary to understand and evaluate the conclusions of this paper are archived at https://doi.org/10.7910/DVN/MYX4CJ. Additional data related to this paper may be requested from the authors.

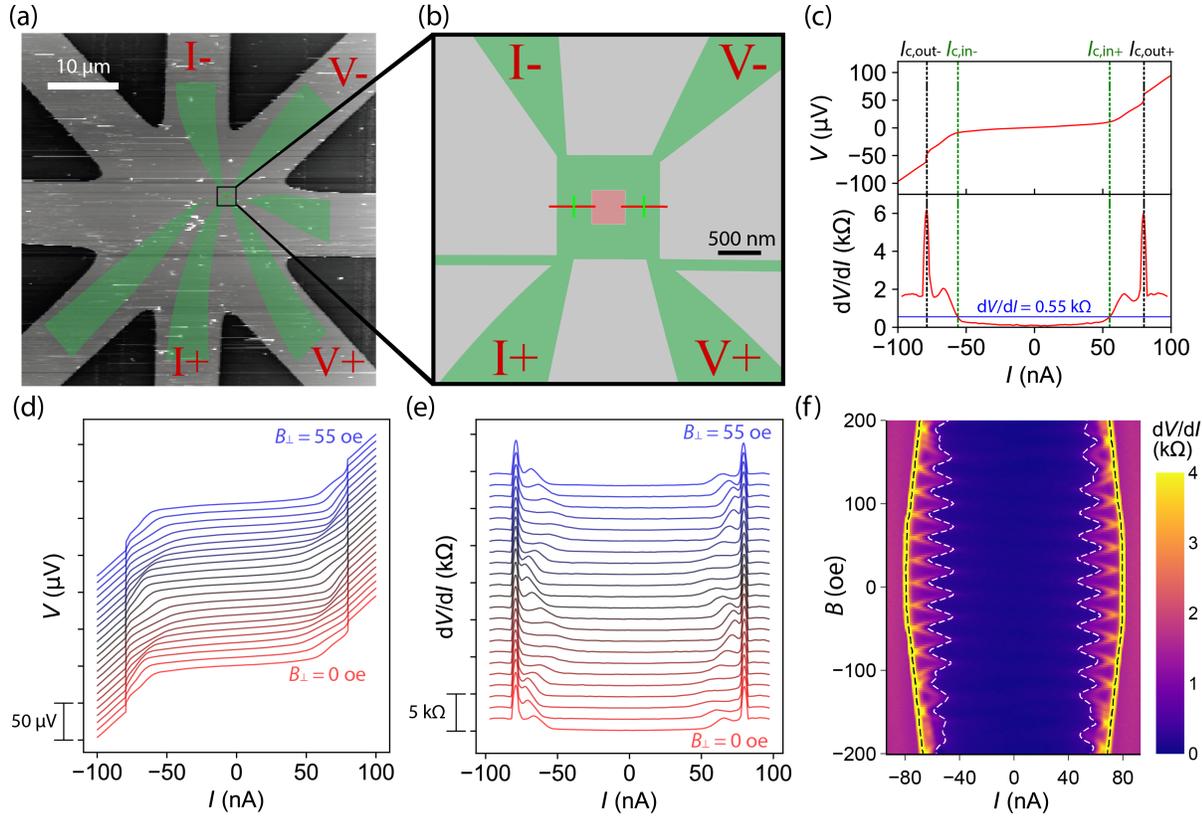

Figure 1. C-AFM lithography and low-temperature current-voltage characteristics of SQUID A. **(a)** AFM height image of the canvas, overlayed by the SQUID design. The green regions are rendered conducting by c-AFM lithography, while red regions are cut (erased) back to insulating (see supplement Fig.S4 for details of the writing process). **(b)** zoom-in view of the central part of Device A, where a 1.2 μm × 1.2 μm sized annulus (inner square size = 0.4 μm × 0.4 μm) is cut into top and bottom half, and then reconnected by the two vertical green nanowires. I+, I-, V+ and V- labels indicate the measurement configurations used for all the following transport experiments. **(c)** Four-terminal $I-V$ characteristics (upper) and corresponding $dV/dI$ vs $I$ curve (lower) measured at $T = 50$ mK and $B = 0$ T. Green and black dashed lines indicate where inner critical current $I_{c,in\pm}$ and outer critical current $I_{c,out\pm}$ are, respectively. $I_{c,in}$ is extracted at the bias current where $dV/dI = R_N/2 = 550\ \Omega$. $I_{c,out}$ is extracted at the coherence peaks where $dV/dI$ reaches maximum. In this case, $I_{c,in+} = 55$ nA, $I_{c,in-} = -56$ nA, $I_{c,out+} = 80$ nA, $I_{c,out-} = -80$ nA. **(d)(e)** $V$ vs $I$ curves and $dV/dI$ vs $I$ curves measured from $B = 0$ Oe to $B = 55$ Oe, with a step of 2.5 Oe. The curves are shifted vertically for clarity. **(f)** Intensity plot of $dV/dI$ vs $I$ and $B$. The white dashed lines mark the value of $I_{c,in\pm}$, while black dashed lines indicate $I_{c,out\pm}$. All measurements in this figure are taken at $T = 50$ mK.

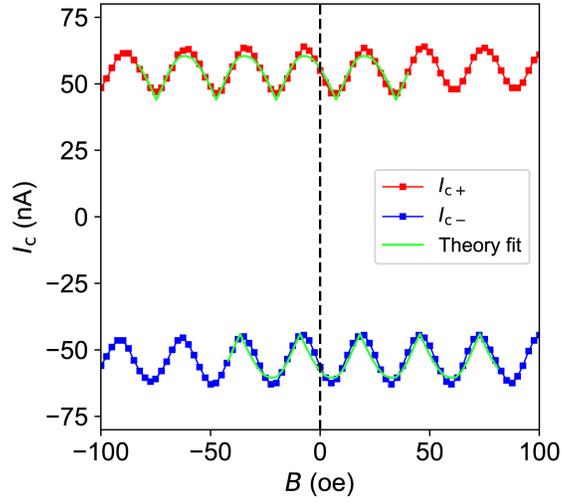

Figure 2. The critical current $I_{c\pm}$ vs $B$ analysis for Device A. Red and blue curves indicate $I_{c+}$ and $I_{c-}$ respectively, replotted from Fig.1(f). The green lines are the theoretical fit using following parameters: $I_{c,l}$ = 26.4 nA, $I_{c,r}$ = 34.0 nA, $\varphi_{c,l}$ = 3.50 rad and $\varphi_{c,r}$ = 4.81 rad.

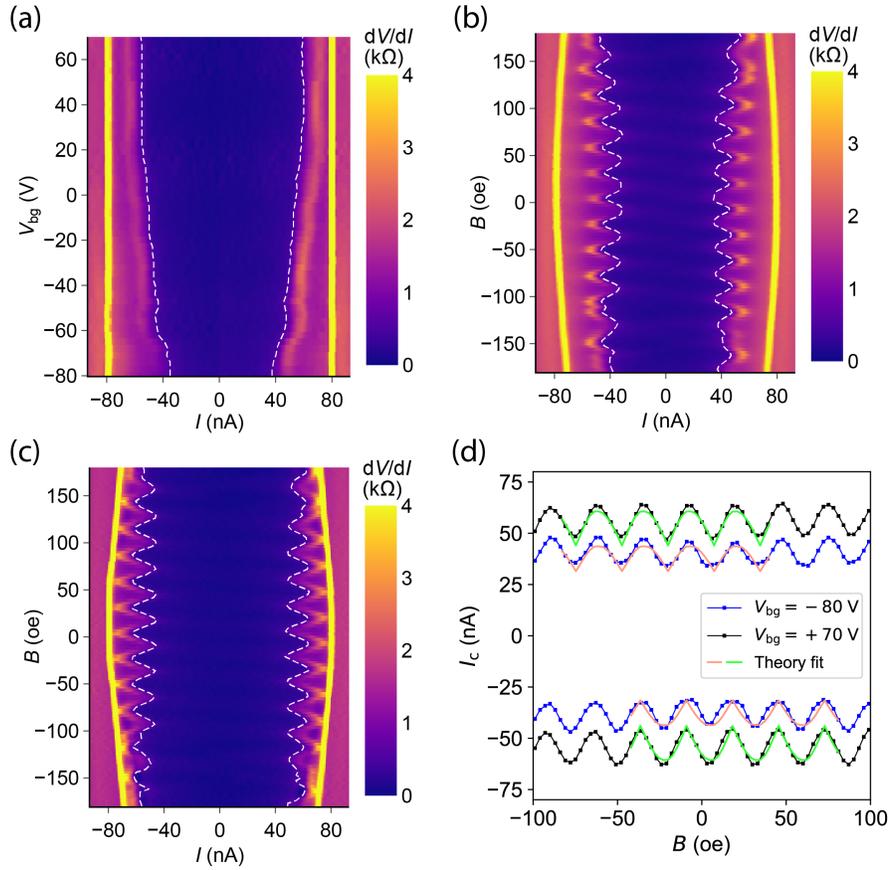

Figure 3. Backgate dependence of SQUID A, taken at $T = 50$ mK. **(a)** Intensity plot of $dV/dI$ vs $I$ and backgate voltage $V_{bg}$. The white dashed lines mark the value of $I_{c\pm}$, where its magnitude decrease with decreasing $V_{bg}$. **(b)(c)** Intensity plot of $dV/dI$ vs $I$ and $B$, measured at $V_{bg} = -80$ V and $V_{bg} = +70$ V respectively. White dashed lines indicate $I_{c\pm}$. **(d)** Critical current $I_{c\pm}$ as a function of magnetic field, comparing between $V_{bg} = -80$ V and $V_{bg} = +70$ V. Orange and green lines are the theoretical fit for $V_{bg} = -80$ V and $V_{bg} = +70$ V respectively, using parameters in Table 2.

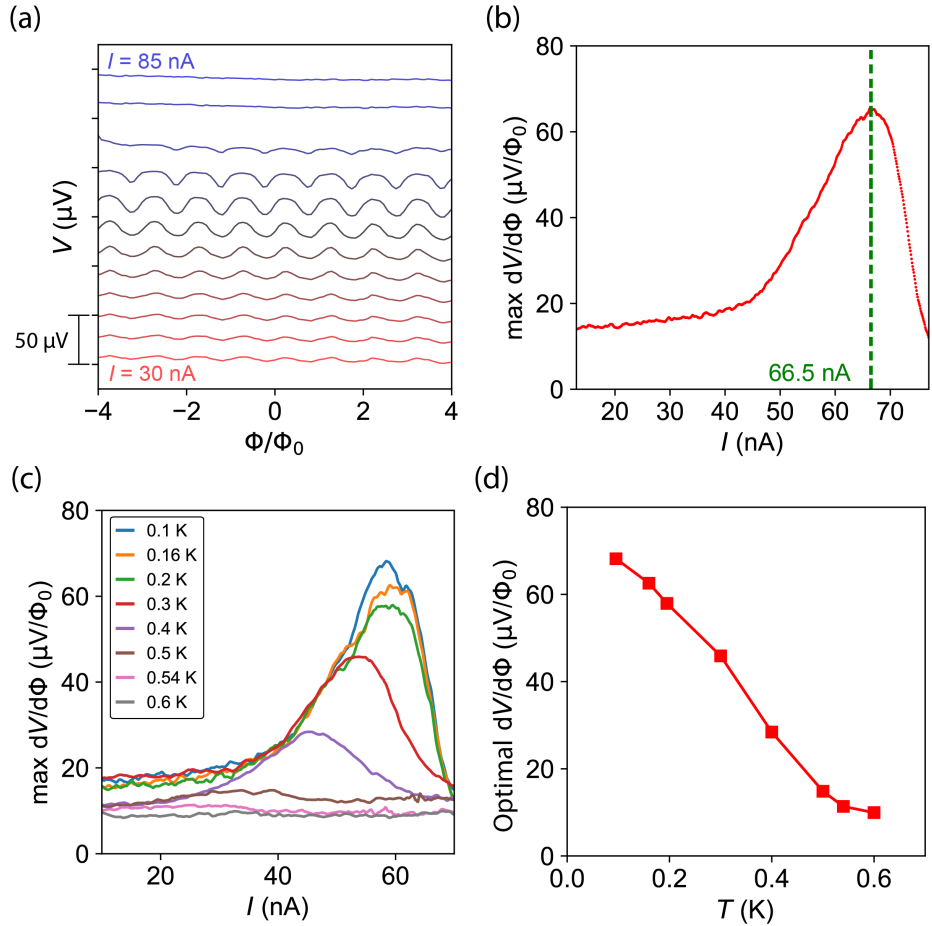

Figure 4. Operating SQUID A as a flux-to-voltage transformer. **(a)** Voltage $V$ modulated by magnetic flux $\Phi$, for different bias currents from 30 nA to 85 nA, in step of 5 nA. The curves are shifted vertically for clarity. This measurement is done at $T$ = 50 mK. **(b)** Corresponding max values of $|dV/d\Phi|$ versus bias current. The optimal bias current for Device A at $T$ = 50 mK is 66.5 nA, where optimal $|dV/d\Phi|$ = 65 μV/$\Phi_0$. **(c)** $|dV/d\Phi|$ maximum versus bias current, measured at elevated temperatures. The optimal bias current decrease with increasing temperature. **(d)** Optimal $|dV/d\Phi|$ extracted at different temperatures from (c).

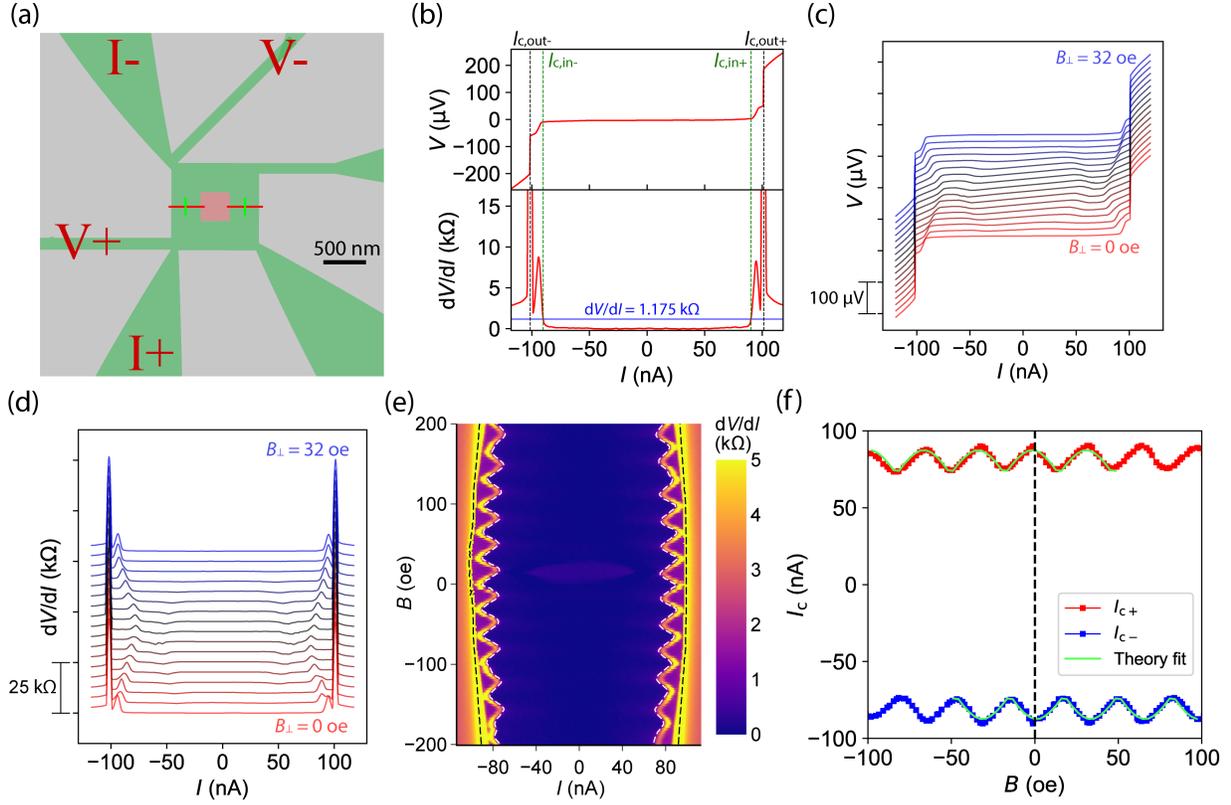

Figure 5. Current-voltage characteristics of SQUID B, taken at $T = 50$ mK. **(a)** C-AFM layout of Device B, which consists of a 1.02 μm × 1.02 μm sized annulus (inner square size = 0.34 μm × 0.34 μm). I+, I-, V+ and V- indicate the measurement configuration. **(b)** $I-V$ characteristics (upper) and corresponding $dV/dI$ vs $I$ curve (lower) measured at $T = 50$ mK and $B = 0$ T. Green and black arrows indicate where inner critical current $I_{c,in\pm}$ and outer critical current $I_{c,out\pm}$ are, respectively. $I_{c,in}$ is extracted at the bias current where $dV/dI = R_N/2 = 1175$ Ω. $I_{c,out}$ is extracted at the coherence peaks where $dV/dI$ reaches maximum. In this case, $I_{c,in+} = 90$ nA, $I_{c,in-} = -90$ nA, $I_{c,out+} = 101$ nA, $I_{c,out-} = -101$ nA. **(c)(d)** $V$ vs $I$ curves and $dV/dI$ vs $I$ curves measured from $B = 0$ Oe to $B = 32$ Oe, with a step of 2 Oe. The curves are shifted vertically for clarity. **(e)** Intensity plot of $dV/dI$ vs $I$ and $B$. The white dashed lines mark the value of $I_{c,in\pm}$, while black dashed lines indicate $I_{c,out\pm}$. **(f)** $I_{c,in\pm}$ as a function of $B$. The green lines are the theoretical fit using following parameters: $I_{c,l} = 40.8$ nA, $I_{c,r} = 46.5$ nA, $\varphi_{c,l} = 5.68$ rad and $\varphi_{c,r} = 5.83$ rad.

# Supplementary Information

# Sketched Nanoscale KTaO$_3$-Based Superconducting Quantum Interference Device

Muqing Yu[1], Nicholas Hougland[1], Qianheng Du[2], Junyi Yang[2], Sayanwitha Biswas[1], Ranjani Ramachandran[1], Dengyu Yang[1,3], Anand Bhattacharya[2], David Pekker[1], Patrick Irvin[1], Jeremy Levy[1,*]

[1]*Department of Physics and Astronomy, University of Pittsburgh*
[2]*Materials Science Division, Argonne National Laboratory*
[3]*Department of Physics, Carnegie Mellon University*
* *jlevy@pitt.edu*

**This supplementary file includes:**

Supplementary Note 1. Voltage Oscillations at zero bias current in SQUID A

Supplementary Note 2. Magnetic Hysteresis in SQUIDs

Figure S1. LAO/KTO sample preparation.

Figure S2. C-AFM erasing and writing of 4nm LAO/KTO (110) sample.

Figure S3. Carrier density and Ginzburg-Landau coherence length characterization of 4nm LAO/KTO (110) sample.

Figure S4. Detailed procedure of creating SQUIDs A and B.

Figure S5. Conductance change of SQUIDs during c-AFM lithography.

Figure S6. Additional information on $I - V$ characteristics of Device A and B.

Figure S7. Temperature dependence of critical current oscillations of SQUID A.

Figure S8. Sensitivity characterization of SQUID B.

Figure S9. $I - V$ characteristics of SQUID B, measured with different configurations, at $T = 50$ mK.

Figure S10. Residual resistance in SQUID A.

Figure S11. Residual resistance in SQUID B and its hysteresis under magnetic field.

Figure S12. Magnetic hysteresis in SQUID A.

Figure S13. SQUID C.

## Supplementary Note 1. Voltage Oscillations at zero bias current in SQUID A

In SQUID A, the periodic voltage oscillations against magnetic field persist at currents below $I_{c,in}$ (main text Fig.4(a)). We note that in SQUID A there is periodic voltage oscillation with magnetic field even at zero bias ($I=0$ nA) (Fig.S17(b)). There can be instrument-related voltage drift over time appearing in the measured $I-V$ curves, as is calibrated by a resistor at room temperature (Fig.S17(c)) and also observed in SQUID B, C and D (Fig.S17(b)). In SQUID A, however, the voltage offset at zero bias is modulated by magnetic field, with the same period as its $I_c$ oscillations. Currently we don't have an explanation for this unphysical behavior, and it is not reproduced in SQUID B, C and D. Capacitive effects are unlikely as each $I-V$ curve takes ~20 s to measure, and there is no visible hysteresis between the forward and reverse current sweeps in SQUID A near zero bias (Fig.S6(a)). The origin might be related to certain special properties of SQUID A, combined with unknown measurement errors. We also note that this issue does not affect our analysis on the critical currents of our devices, which are extracted from differential resistance $dV/dI$ with no dependence on the voltage offset of $I-V$ curves.

## Supplementary Note 2. Magnetic Hysteresis in SQUIDs

Magnetic hysteresis in SQUID B can be inferred from its residual as a function of magnetic field. After SQUID B is patterned by c-AFM lithography and cooled down from room temperature (Fig.S11(a)), the very first $B$ field sweep is performed from +250 oe to -250 oe while measuring $I-V$ curves. A resistive feature at $0\text{ oe}<B<+30\text{ oe}$ and $-40\text{ nA}<I<+40\text{ nA}$ is revealed (Fig.S11(b)). The residual resistance stays at $0\text{ oe}<B<+30\text{ oe}$ in subsequent small-ranged $B$ field sweeps between $\pm250$ oe. $B$ field is then ramped to $-4000$ oe and brought back to near 0 oe, after which the resistive feature shift to $-20\text{ oe}<B<0\text{ oe}$ and $-40\text{ nA}<I<+40\text{ nA}$ (Fig.S11(c)). Fig.S11(d) plots the zero-bias resistance versus $B$ field in these two $B$ sweeps, before and after the $B$ ramping to -4000 oe. The shift in the residual resistance peak is 26.0 oe. We also note that the width of these two resistance peaks are different: the first one spans from 0 oe to +30 oe while the second one spans from -20 oe to 0 oe. The $I_c$ vs $B$ oscillations of SQUID B also shift by 26 oe between these two $B$ sweeps (Fig.S11(e)). The $I_c$ vs $B$ oscillations of SQUID A also exhibit a hysteresis under *B* field before and after *B* ramping to –4000 oe (Fig.S12), although the magnitude of the hysteresis (22.7 oe) is slightly smaller than what is observed in SQUID B. The origin of this hysteresis can be magnetism in KTO [47,48], or trapped vortices in the two half-annuli [54].

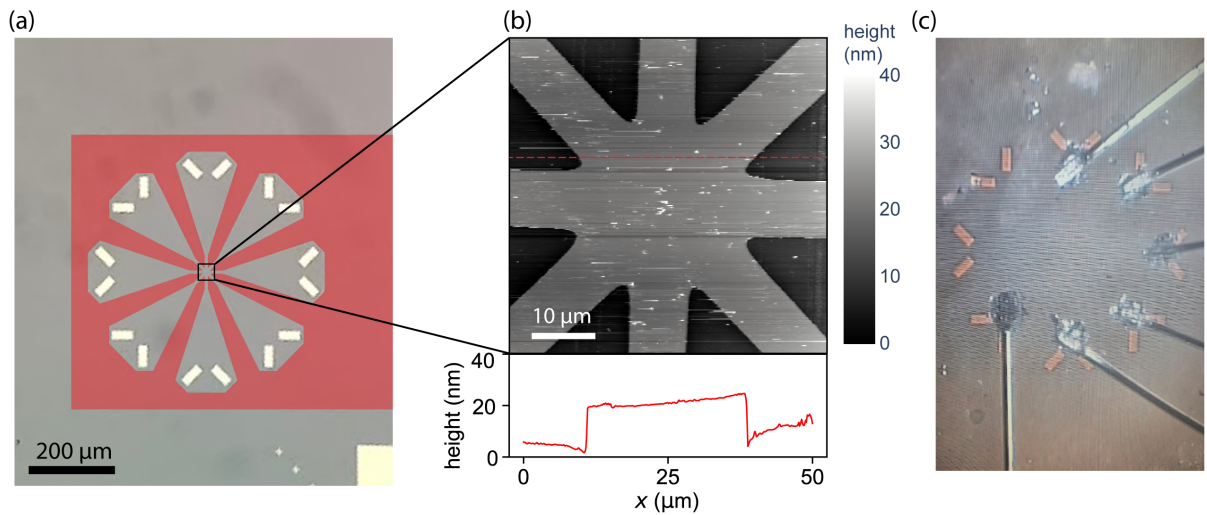

Figure S1. LAO/KTO sample preparation. **(a)** Optical image of the 4nm LAO/KTO (110) sample. In the region that is shaded by red color, LAO is dry etched by ICPRIE (see methods section). **(b)** AFM height image of the c-AFM lithography canvas, which is the central part of (a). The linecut along the red dashed line shows the ICPRIE depth is ~ 15 nm. **(c)** Aluminum wirebonds are used to contact LAO/KTO interface for electrical transport measurements.

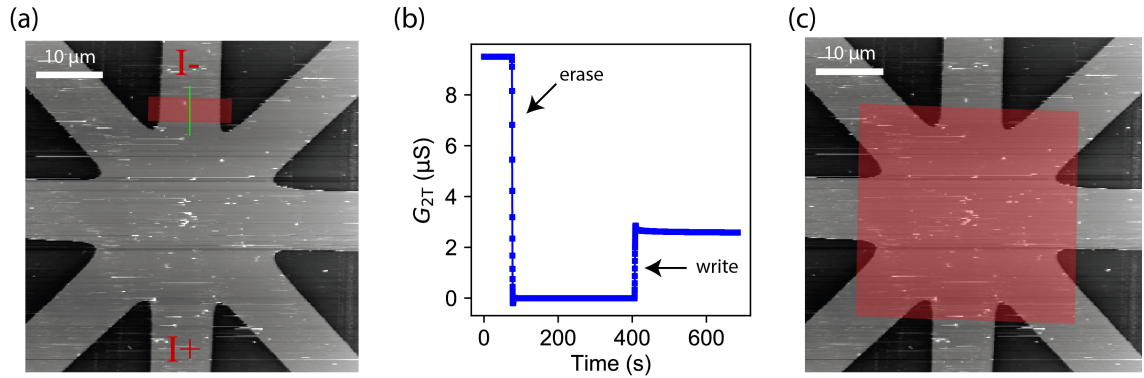

Figure S2. C-AFM erasing and writing of 4nm LAO/KTO (110) sample. **(a)** c-AFM lithography pattern overlaying AFM height image. The AFM tip first renders the red rectangle with $V_{\text{tip}} = -12$ V (erasing), and then sketches along the green wire with $V_{\text{tip}} = +20$ V (writing). **(b)** Before any c-AFM lithography, the intrinsic two-terminal conductance ($G_{2T}$) between I+ and I- is ~9 µs. $G_{2T}$ drops to 0 during erasing and then recovers during the writing of the nanowire. **(c)** The whole canvas is restored to insulating by c-AFM erasing with the big red rectangle, prior to writing each device described in this manuscript.

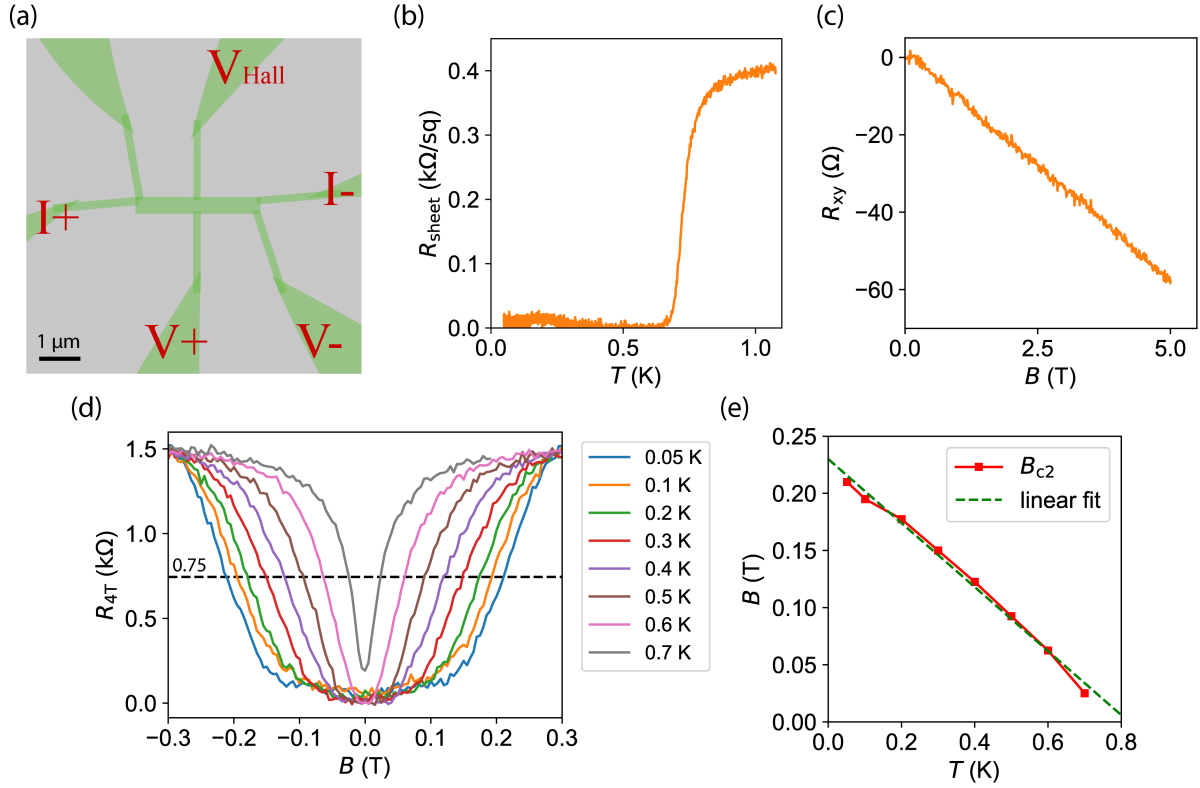

Figure S3. Carrier density and Ginzburg-Landau coherence length characterization of 4 nm LAO/KTO (110) sample. **(a)** C-AFM layout of the Hallbar device. Main channel is written with $V_{\text{tip}} = +33$ V. The width $w = 0.4$ μm, and its aspect ratio is 3.6. **(b)** Sheet resistance $R_{\text{sheet}}$ versus temperature $T$. **(c)** Hall resistance $R_{\text{xy}}$ versus magnetic field $B$, measured at $T = 1.1$ K. The carrier density can be extracted: $n_{2D} = 5.4 \times 10^{13}$ cm$^{-2}$. At $T = 1.1$ K, Mobility $\mu = 1/n_{2D}eR_{\text{sheet}} = 280$ cm$^2$/V/s. We can also get mean free path of the electrons: $l_{mfp} = v_F\tau = \mu\hbar\sqrt{2\pi n_{2D}}/e = 33$ nm. **(d)** Four-terminal resistance $R_{4T}$ measured at different temperatures as a function of $B$, from $T = 50$ mK to $T = 700$ mK. The back dashed line indicates 0.75 kΩ, which is half of the normal state resistance. **(e)** Temperature dependence of upper critical field, extracted at $R_{4T} = 0.75$ kΩ from (d). Green dashed line is the linear fit with: $B_{C2} = \Phi_0/2\pi\xi_{GL}^2 \times (1 - T/T_C)$, which gives Ginzburg-Landau coherence length: $\xi_{GL} = 38$ nm.

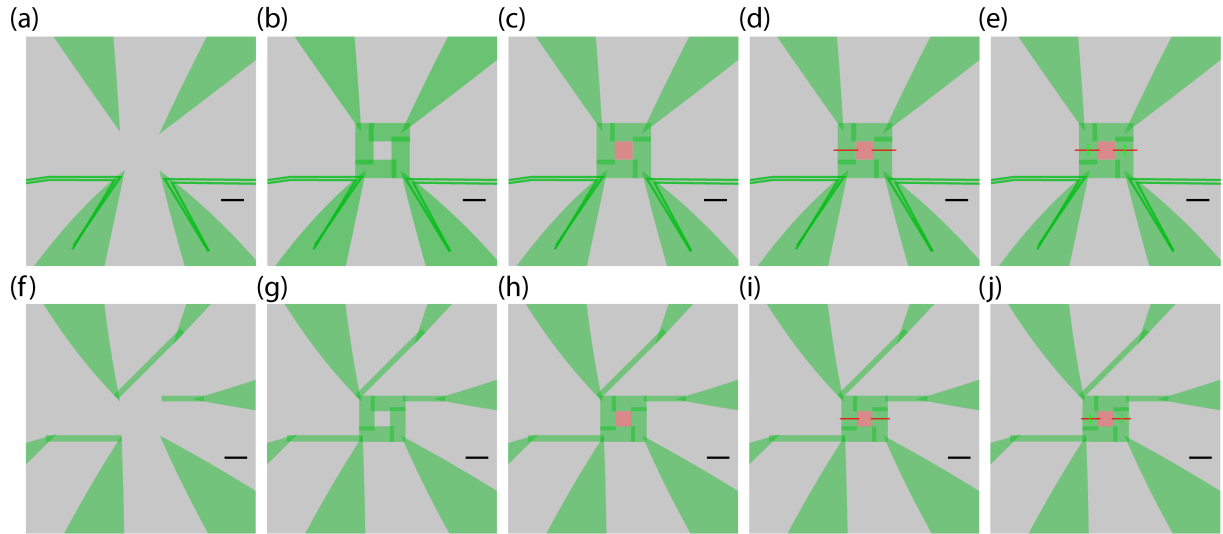

Figure S4. Detailed procedure of creating SQUID A and B. **(a)-(e)** Creation of Device A. **(a)** Write current and voltage leads. Tip voltage used: $V_{tip} = +37$ V. **(b)** Write the annulus, which consists of four rectangles, with $V_{tip} = +33$ V. We note the overlapping part of the rectangles are written twice, which could result in a higher carrier density in that region and inhomogeneous superconductivity across the annulus. **(c)** Erase in the center hole with $V_{tip} = -10$ V, to make sure it is insulating. **(d)** Cut the annulus to top and bottom half, by sketching across each arm once with $V_{tip} = -12$ V. **(e)** Reconnect top and bottom half by sketching along the light green paths once, with $V_{tip} = +8$ V. These two nanowires serve as the two WLs for this SQUID. **(f-j)** Creation of Device B. The procedure is identical except the two WLs are sketched with $V_{tip} = +9$ V in (j). Same as Device A, we note the overlapping of rectangles when writing the annulus of Device B, which can lead to inhomogeneous superconductivity. The scalebars in all images are 500 nm.

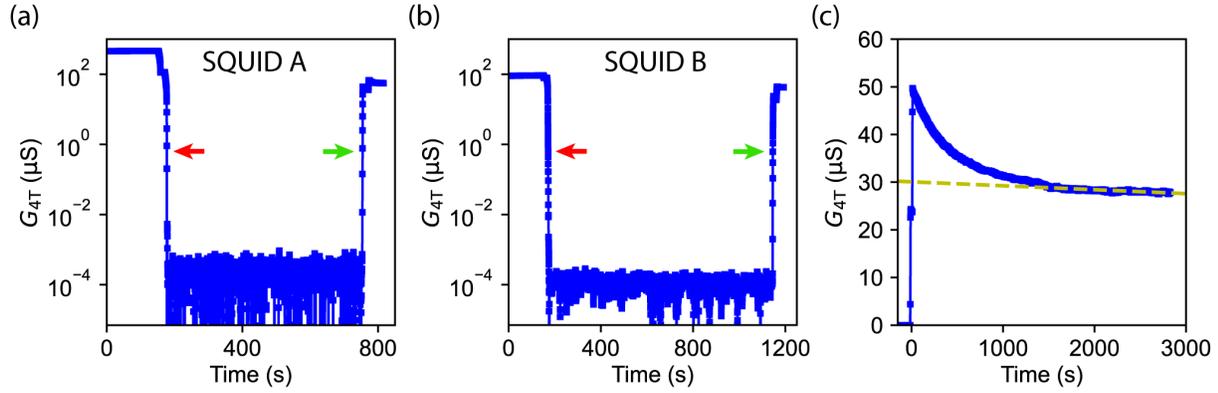

Figure S5. Conductance change of SQUIDs during c-AFM lithography. **(a)(b)** Four-terminal conductance $G_{4T}$ monitored as a function of time during the c-AFM lithography of Device A(B). The four-terminal measurement configurations are shown in main text Fig.1(b) and Fig.5(a) respectively. Red arrows indicate the conductance drops when the annulus is cut into top and bottom half (Fig.S4(d)(i)). $G_{4T}$ stays near 0 (below the noise floor) until WLs are written (Fig.S4(e)(j)), which gives the conductance jumps indicated by green arrows. **(c)** To calibrate the lifetime of our SQUIDs in ambient conditions (27 °C, 40% humidity), a device is created with identical layout and c-AFM parameters as SQUID A, and then its $G_{4T}$ is continuously monitored for ~50 min. After the two WLs are written, $G_{4T}$ decays exponentially in the first 1500 s and then stabilizes into a slower linear decay, which can be fitted using $G_{4T}(t) = 30~\mu S - 0.8~\text{nS/s} \times t$. Extrapolating this linear fit to 0 μS gives a lifetime estimate of our SQUIDs ~10 h in ambient conditions.

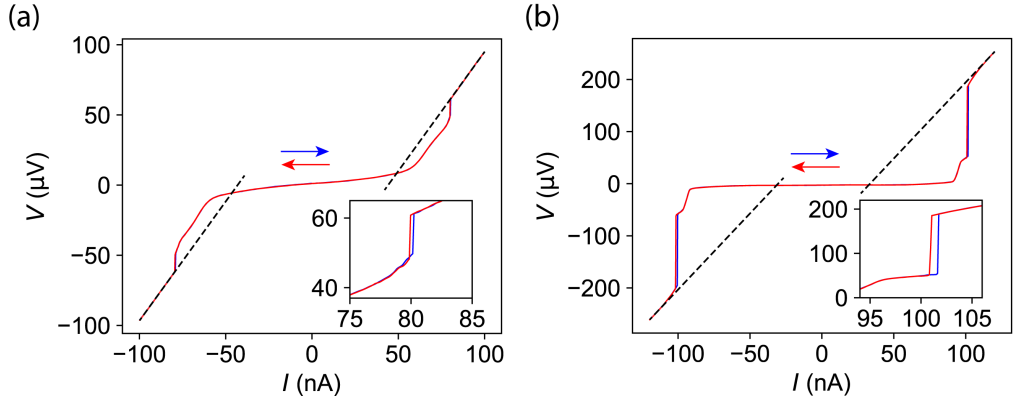

Figure S6. Additional information on $I-V$ characteristics of Device A and B. **(a)(b)** $I-V$ curve of Device A(B), measured at $T = 50$ mK and $B = 0$ T. Blue and red arrows indicate the current sweep direction. Black dashed lines: linear extension of the $I > I_{c,out+}$ portion of the $I-V$ curve. Inset: zoom-in view near $I_{c,out+}$ of Device A(B), where a 0.4 nA (1 nA) hysteresis is observed forward and backward sweeps. Because the relatively small hysteresis compared to the magnitude of critical current ($< 1\%$), in other figures, the red curves (backward sweep) are chosen to represent the $I-V$ curve of Device A(B).

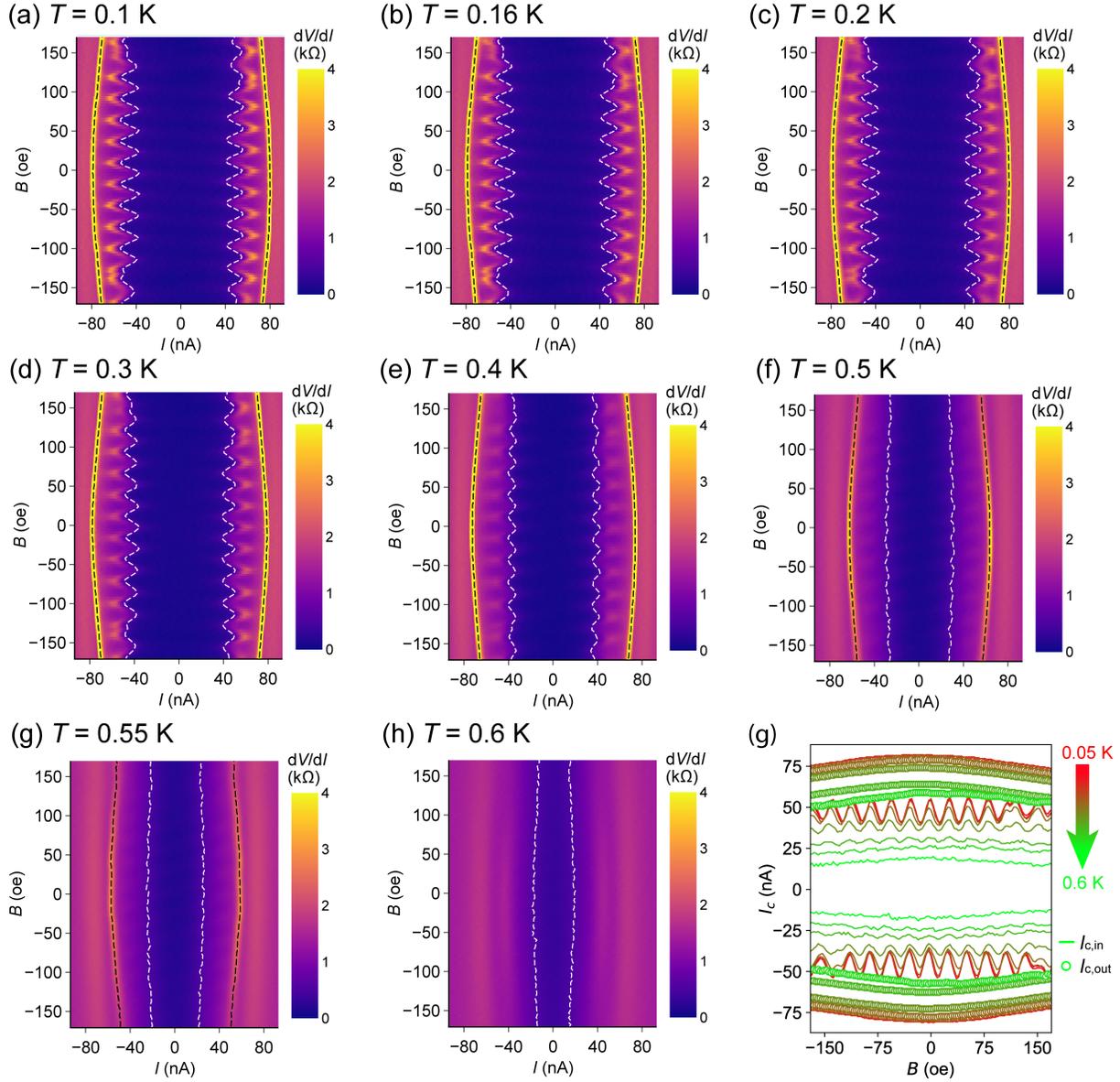

Figure S7. Temperature dependence of critical current oscillations of SQUID A. **(a)-(h)** Intensity plot of $dV/dI$ vs $I$ and $B$, measured at different temperatures ranging from 0.1 K to 0.6 K. White dashed lines indicate $I_{c,in\pm}$ and black dashed lines indicate $I_{c,out\pm}$. We note that $I_{c,out\pm}$ at $T = 0.6$ K can not be extracted from $dV/dI$ peaks and thus is not plotted. **(g)** $I_{c,in\pm}$ and $I_{c,out\pm}$ from (a)-(h) are replotted, as a function of magnetic field, at different temperatures. The periodic oscillation of $I_{c,in\pm}$ vanishes at temperature $T \geq 0.6$ K.

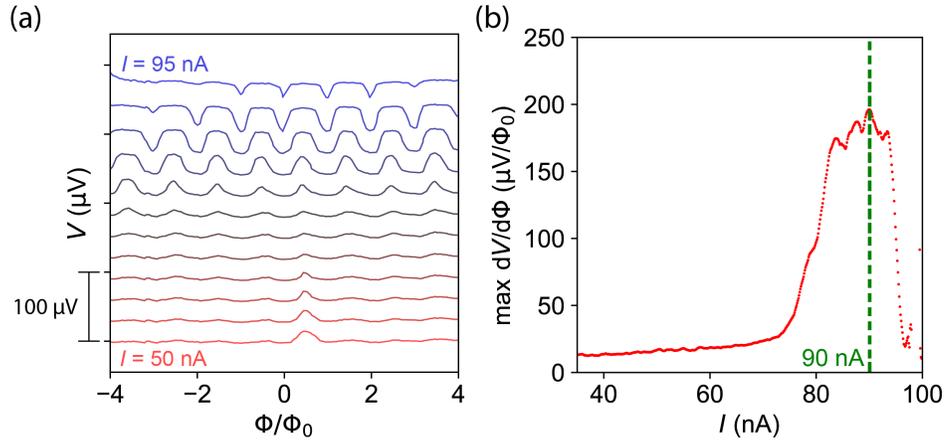

Figure S8. Sensitivity characterization of SQUID B. **(a)** Voltage $V$ modulated by magnetic flux $\Phi$, for different bias currents from 50 nA to 94 nA, in step of 4 nA. The curves are shifted vertically for clarity. This measurement is done at $T$ = 50 mK. **(b)** Corresponding max values of $|dV/d\Phi|$ versus bias current. The optimal bias current is 90 nA, where optimal $|dV/d\Phi|$ = 190 µV/$\Phi_0$.

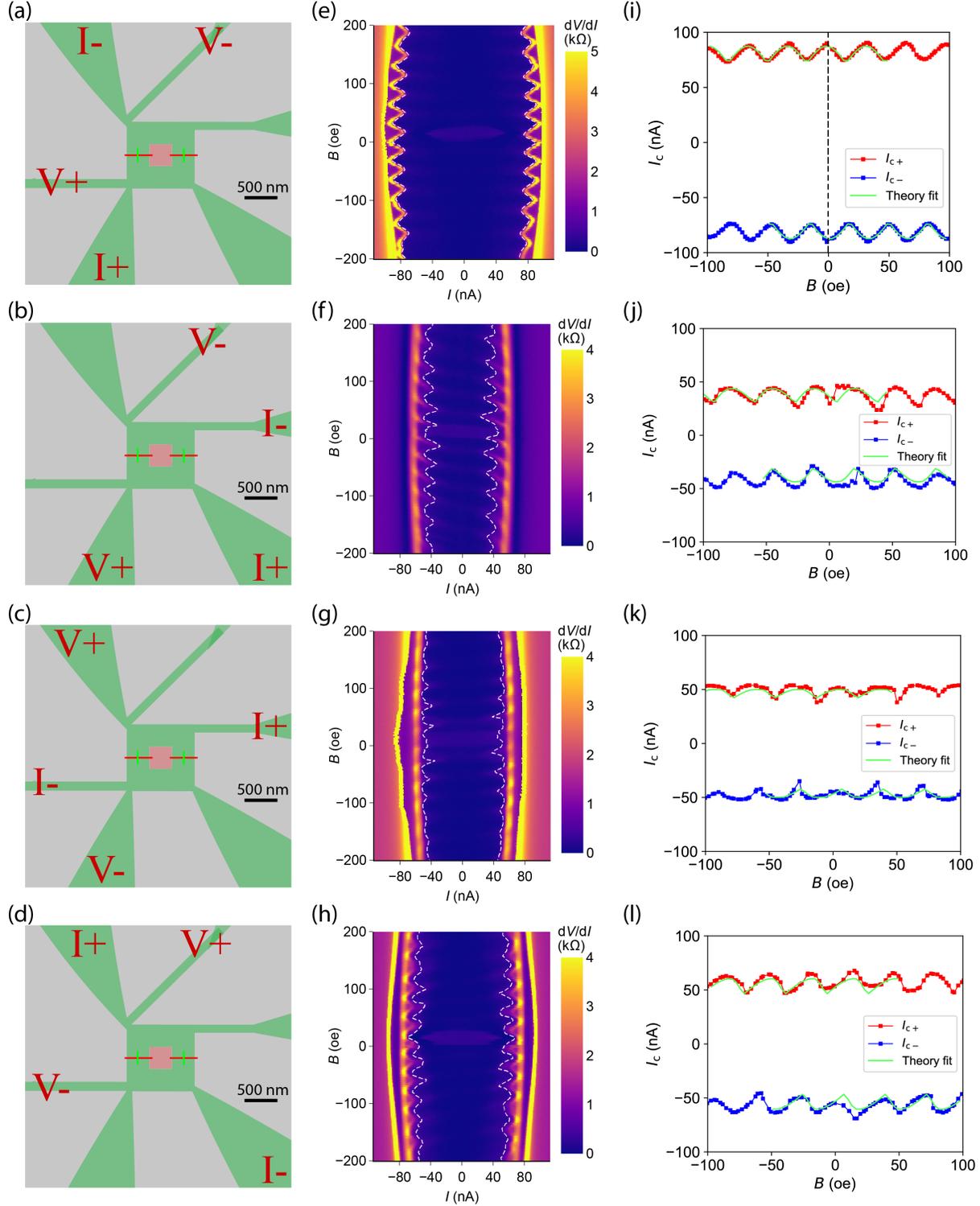

Figure S9. $I-V$ characteristics of SQUID B, measured with different configurations, at $T=50$ mK. **(a)(b)(c)(d)** Measurement configurations **(e)(f)(g)(h)** Intensity plot of d$V$/d$I$ vs $I$ and $B$, corresponding with (a)-(d). White dashes lines mark the value of $I_{c\pm}$. **(i)(j)(k)(l)** $I_{c\pm}$ as a function of $B$, overlayed by theoretical fits using parameters in Table 3, corresponding with (a)-(d).

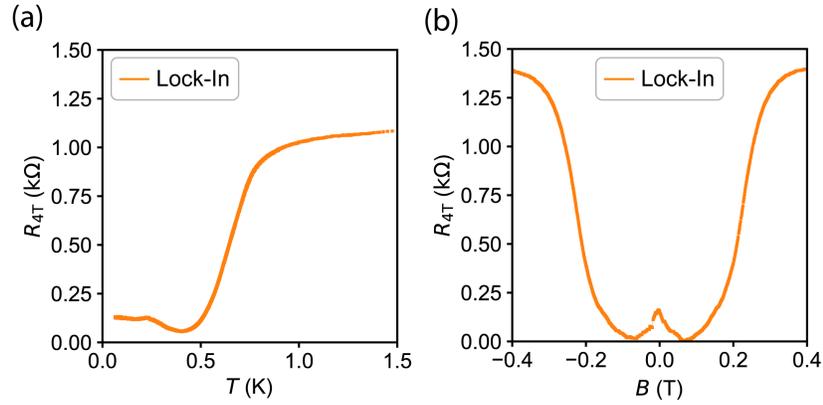

Figure S10. Residual resistance in SQUID A. **(a)** Lock-in measurement of Device A resistance as a function of temperature at $B = 0$ T. At $T = 1.5$ K, we can extract the normal state resistance $R_\text{n} = 1.1$ kΩ. The superconducting transition spans approximately from 1 K to 0.5 K. An increase in resistance can be observed as temperature further lowers below 0.4 K. The resistance subsequently plateaus below 0.2 K. **(b)** Lock-in measurement of Device A resistance as a function of magnetic field near $T = 50$ mK. A resistance peak that spans from $-0.07$ T to $0.07$ T is observed.

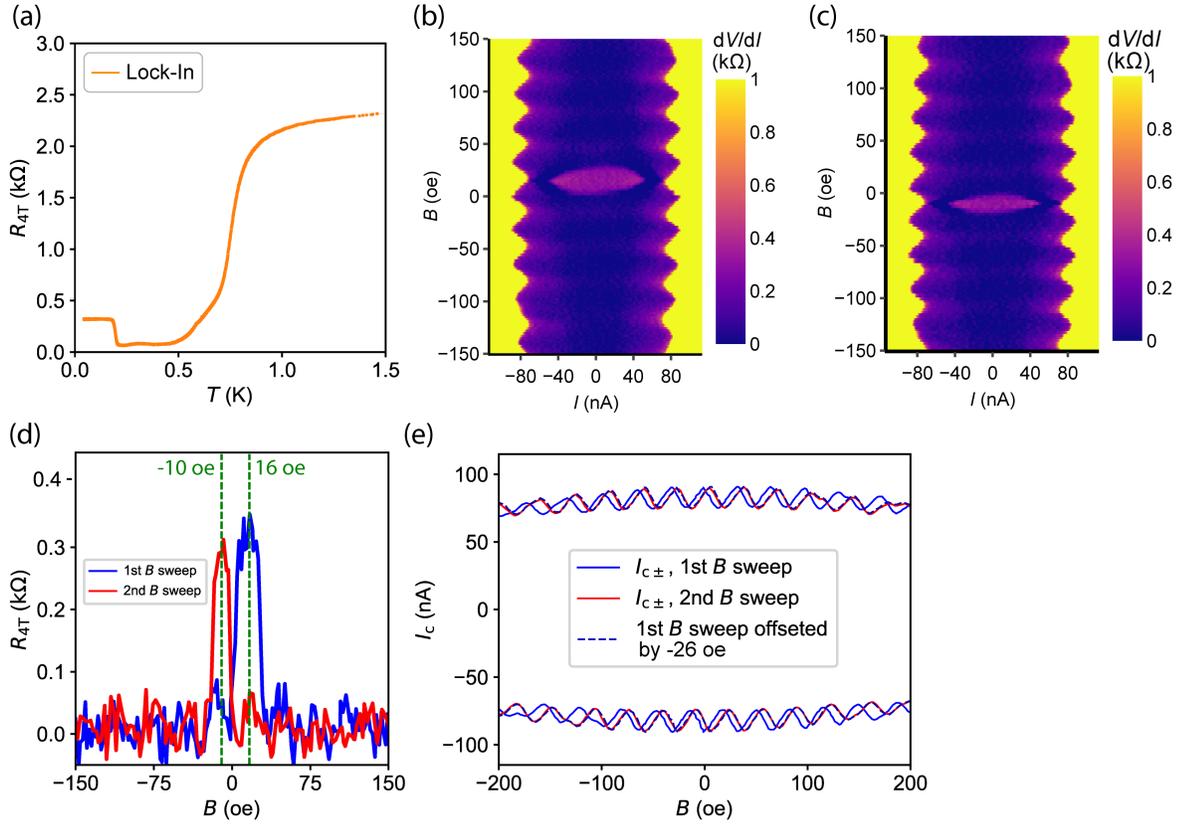

Figure S11. Residual resistance in SQUID B and its hysteresis under magnetic field. **(a)** Lock-in measurement of resistance of Device B versus temperature at $B = 0$ T. This measurement is taken before any other measurements that involve sweeping $B$. At $T = 1.5$ K, its normal state resistance is $R_n = 2.35$ kΩ. As temperature decreases, superconducting transition of the device is followed by a sharp increase in its resistance near 0.2 K, which plateaus at the lowest temperature. **(b)** Intensity plot of $dV/dI$ vs $I$ and $B$, replotted from main text Fig.5(e) with a different color scale. Prior to this measurement, $B$ is ramped from 0 to $+250$ oe. An obvious finite $dV/dI$ regime is observed when 0 oe $< B < +30$ oe and $-40$ nA $< I < +40$ nA. This resistive feature remains at 0 oe $< B < +30$ oe when the subsequent $B$ field sweeps were performed between $\pm 250$ oe. **(c)** Intensity plot of $dV/dI$ vs $I$ and $B$, measured after $B$ ramps down to $-4000$ oe and then returns to $-200$ oe. The resistive feature shifted to $-20$ oe $< B <$ 0 oe and $-40$ nA $< I < +40$ nA. **(d)** Comparison between the zero-bias $dV/dI$ vs $B$ relationships of the $1^{st}$ $B$ sweep in (b) and of the $2^{nd}$ $B$ sweep in (c). The $dV/dI$ peak occurs at $+16$ oe and $-10$ oe respectively in these two measurements. $B$ ramping to $-4000$ oe causes the resistance peak to shift by $-26$ oe. **(e)** Comparison between the $I_c$ vs $B$ relationships of the $1^{st}$ $B$ sweep in (b) and of the $2^{nd}$ $B$ sweep in (c). The $I_c$ oscillations are also shifted by $-26.0$ oe between the two measurements.

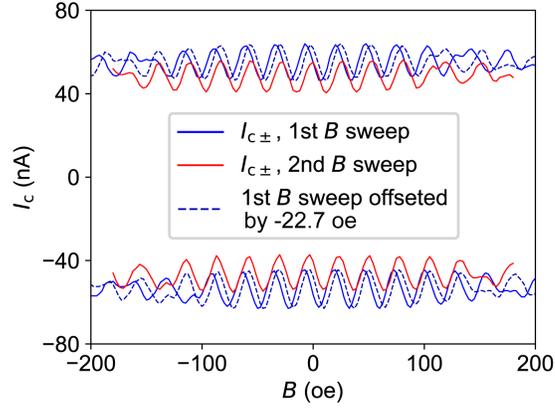

Figure S12. Magnetic hysteresis in SQUID A. $I_c$ vs $B$ relationships are replotted from main text Fig.2 (labeled as the 1st $B$ sweep) and also replotted from Fig.S7(a) (labeled as the 2nd $B$ sweep). Between these two measurements, $B$ is ramped to $-4000$ oe, which causes the magnetic hysteresis in $I_c$ vs $B$. Backgate ($V_{bg}$) was also applied between these two measurements, which causes mismatch in the amplitude of $I_c$. By comparing similar features in $I_c$ vs $B$, we determine $I_c$ in the 2nd $B$ sweep is shifted by $-22.7$ oe from the 1st sweep.

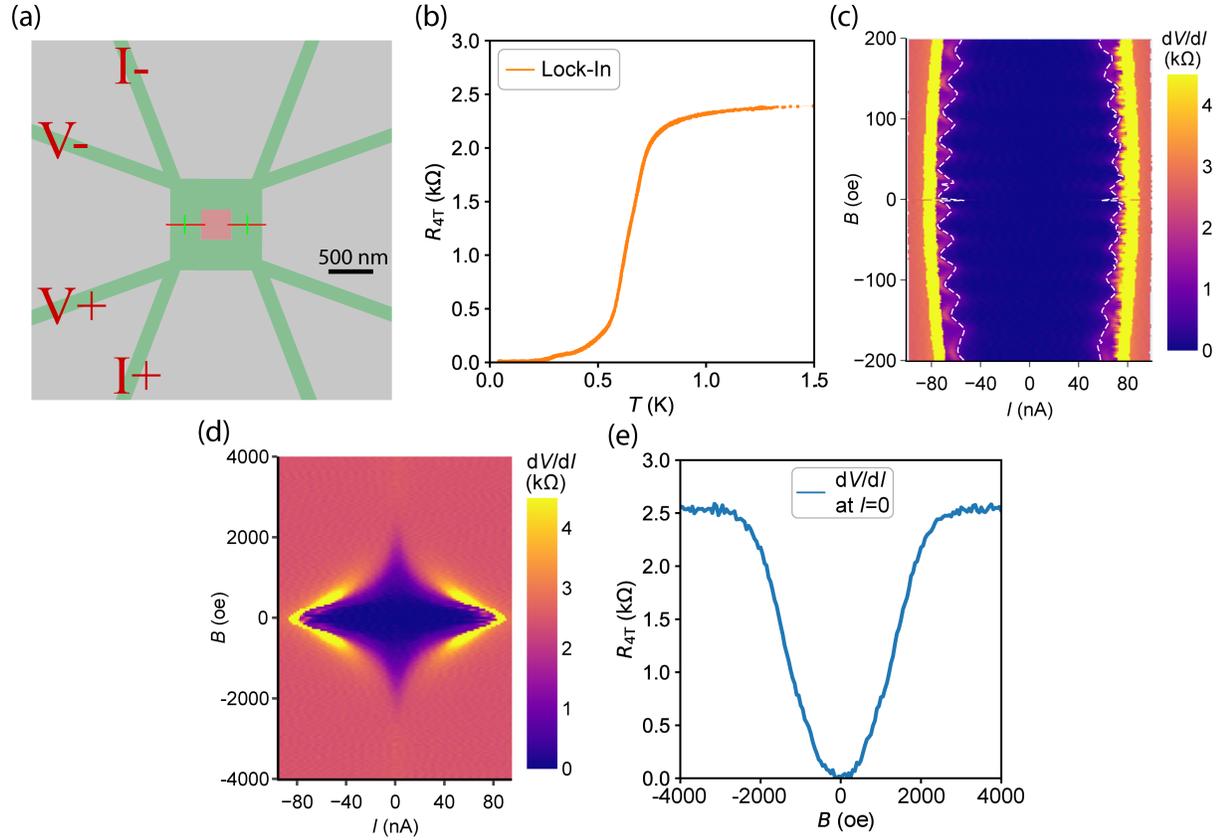

Figure S13. SQUID C. **(a)** C-AFM layout. Two vertical bridges connect the top and bottom half of a conducting annulus. Outer area $A_\text{O} = 1.02$ μm $\times\ 1.02$ μm $= 1.04$ μm$^2$ and inner area (red rectangle) $A_\text{I} = 0.34$ μm $\times\ 0.34$ μm $= 0.12$ μm$^2$. **(b)** Lock-in measurement of resistance $R$ versus temperature at $B = 0$ T. In Device C, $R$ decreases monotonically from $T = 1.5$ K to base temperature $T = 0.05$ K. At $T = 0.05$ K, $R$ is immeasurably small. **(c)** Intensity plot of $dV/dI$ vs $I$ and $B$. The white dashed lines mark the value of $I_{c\pm}$, which have an oscillation period $\Delta B = 36$ Oe. **(d)** Intensity plot of $dV/dI$ vs $I$ and $B$ measured within a larger $B$ range at $T = 0.05$ K. **(e)** Zero-bias $dV/dI$ versus $B$, where $dV/dI$ increases monotonically as $|B|$ increases.

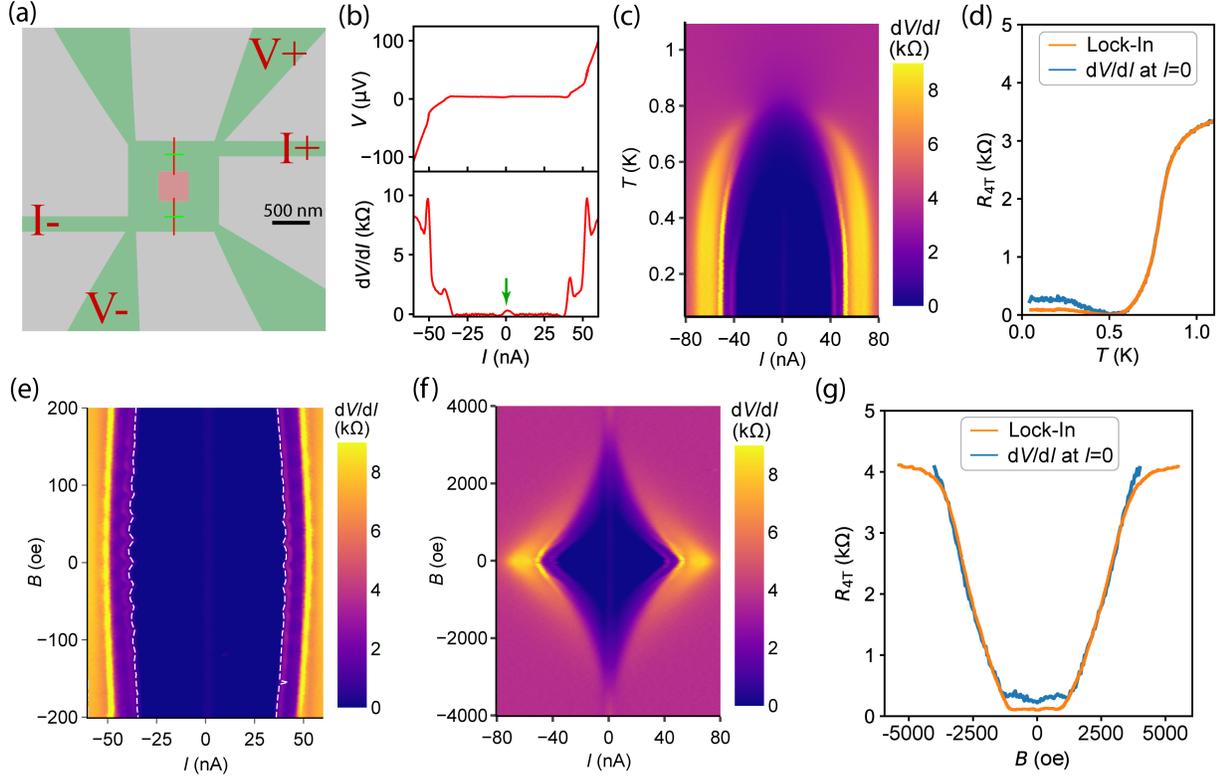

Figure S14. SQUID D. **(a)** C-AFM layout. Two horizontal bridges connect the left and right half of a conducting annulus. Outer area $A_O = 1.2\ \mu m \times 1.2\ \mu m = 1.44\ \mu m^2$ and inner area (red rectangle) $A_I = 0.4\ \mu m \times 0.4\ \mu m = 0.16\ \mu m^2$. **(b)** $I - V$ characteristics (upper) and corresponding $dV/dI$ vs $I$ curve (lower) measured at $T = 50$ mK and $B = 0$ T, where a resistance peak near zero-bias is observed and marked by a green arrow. **(c)** Intensity plot of $dV/dI$ vs $I$ and $T$, measured at $B = 0$ T. **(d)** Zero-bias $dV/dI$ (blue) and also lock-in measurement of resistance (orange) versus temperature at $B = 0$ T. The relatively large ac amplitude of current used in lock-in measurements (see Methods section) results in its failure to capture the zero-bias resistance in (b), thus its deviation from zero-bias $dV/dI$. In both measurements, as $T$ decreases from 1 K, $R$ decreases to close to 0 near 0.5 K and then increases and eventually plateaus at $T < 0.25$ K. **(e)** Intensity plot of $dV/dI$ vs $I$ and $B$. The white dashed lines mark the value of $I_{c\pm}$, which have an oscillation period $\Delta B = 29$ Oe. The small amplitude of $I_c$ modulation is likely due to the large mismatch between the critical currents of the two bridges. **(f)** Intensity plot of $dV/dI$ vs $I$ and $B$ measured within a larger $B$ range at $T = 0.05$ K. **(g)** Zero-bias $dV/dI$ and also lock-in measured $R$ versus $B$. The mismatch between them again results from the relatively large ac current used by lock-in. $R$ remains finite near 0 oe, and increases monotonically as $|B|$ increases to higher values.

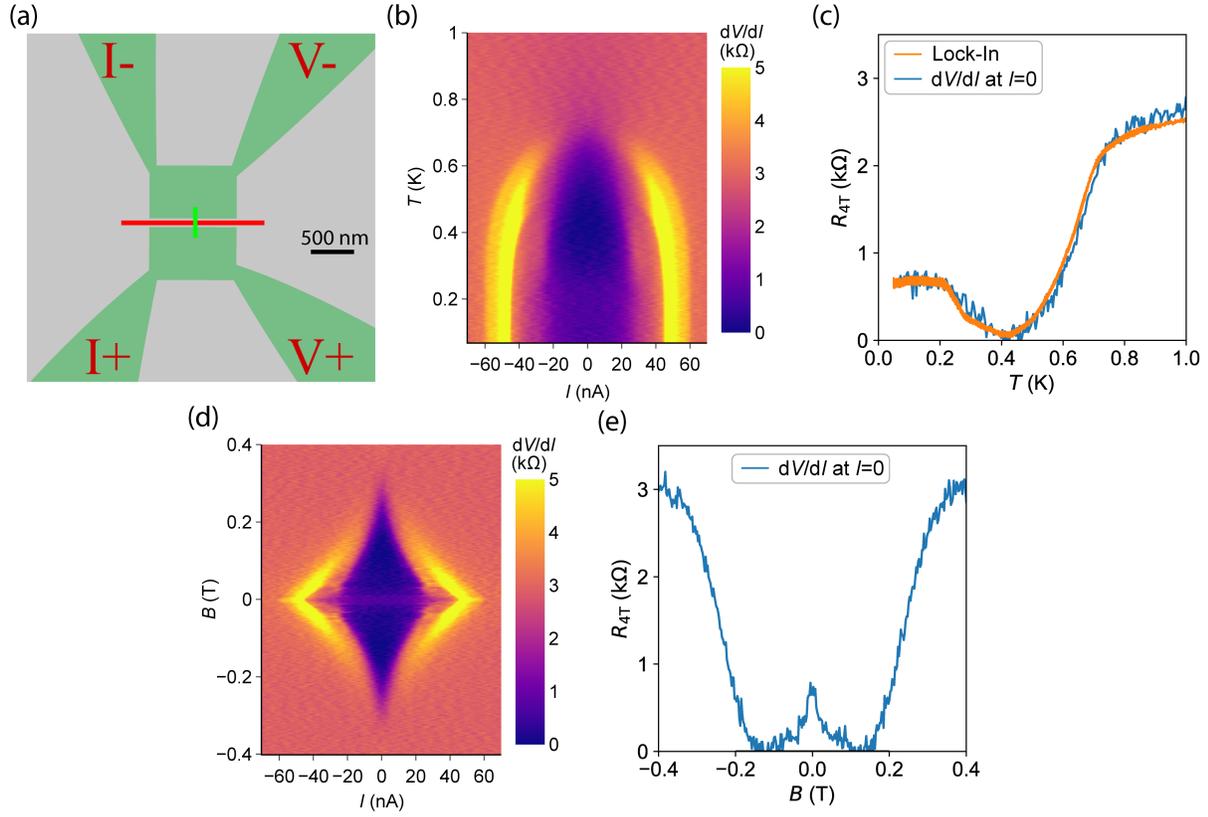

Figure S15. Single Dayem bridge Device E. **(a)** C-AFM layout. A vertical bridge is connecting two rectangular conducting regions. **(b)** Intensity plot of $dV/dI$ vs $I$ and $T$, measured at $B = 0$ T. **(c)** Zero-bias $dV/dI$ and also lock-in measurement of resistance as a function of temperature at $B = 0$ T. As $T$ decreases from 1 K to 0.05 K, $R$ drops to close to 0 kΩ at 0.4 K and starts to increase below 0.4 K, and then plateaus at $T < 0.2$ K. This behavior is similar to the $R$ vs $T$ relationships observed in SQUID A, B and D (Fig.S10(a), Fig.S11(a) and Fig.S14(d)). **(d)** Intensity plot of $dV/dI$ vs $I$ and $B$, measured at $T = 50$ mK. **(e)** Zero-bias $dV/dI$ as a function of magnetic field. A resistance peak near zero field is visible, similar to the $R$ vs $B$ relationship of SQUID A (Fig.S10(b)).

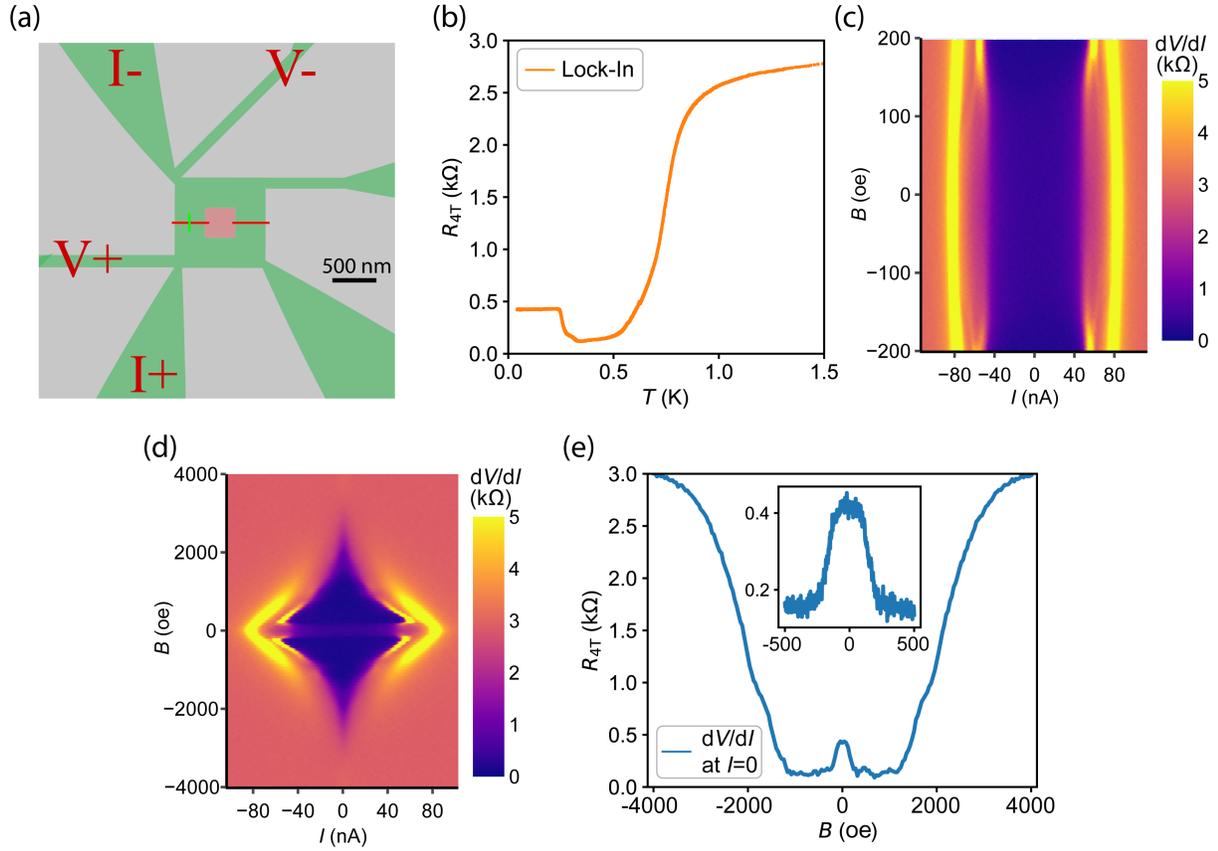

Figure S16. Single Dayem bridge Device F. **(a)** C-AFM layout. The pattern exactly the same as SQUID B (main text Fig.5a) except only the left bridge is written. **(b)** Lock-in measurement of resistance as a function of temperature at $B = 0$ T. A non-monotonic dependence of $R$ on $T$ is observed, similar to the $R$ vs $T$ relationships in Device A, B, D and E (Fig.S10(a), Fig.S11(a), Fig.S14(d), Fig.S15(c)). **(c)** Intensity plot of $dV/dI$ vs $I$ and $B$, measured at $T = 50$ mK. No periodic oscillations of $I - V$ against $B$ can be seen, different from SQUID B (main text Fig.5(e)). **(d)** Intensity plot of $dV/dI$ vs $I$ and $B$, measured within a larger range of magnetic field. **(e)** Zero-bias $dV/dI$ as a function of magnetic field. Inset: zoom-in on the resistance peak near zero field, which spans between $\pm 300$ oe.

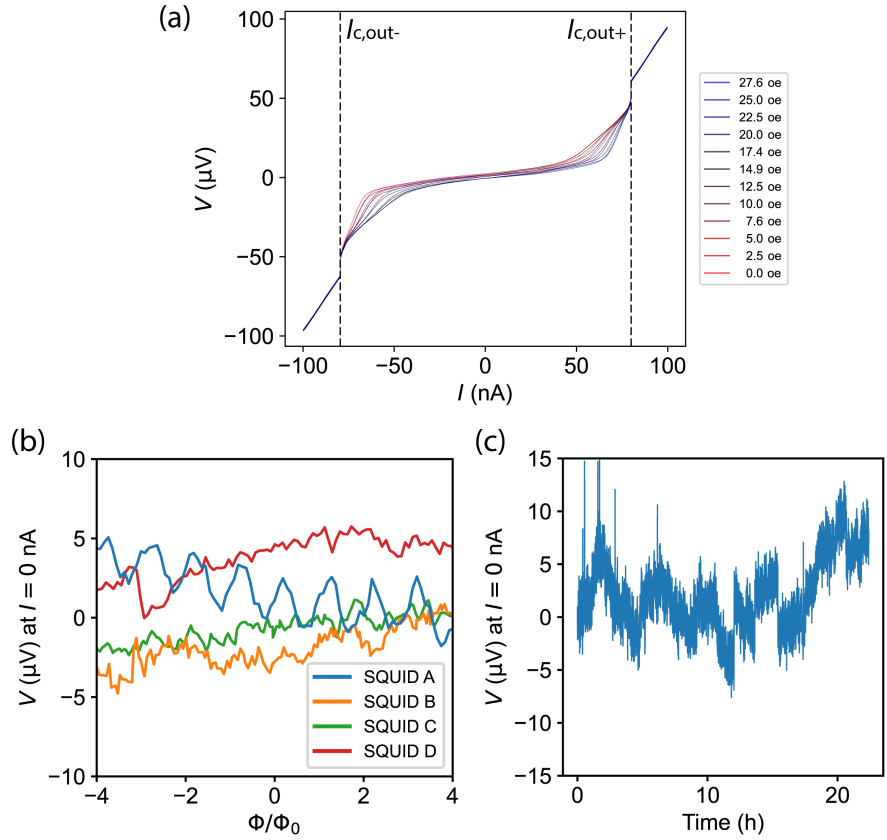

Figure S17. Characterization of voltage drift during $I-V$ measurements. **(a)** $I-V$ curves of SQUID A measured at different magnetic fields from 0 oe to +27.5 oe. Black dashed lines mark the outer critical current $I_{c,out}$. The $I-V$ curves are almost aligned when SQUID A is in the normal regime, i.e. $I > I_{c,out}$. Within the superconducting regime, $I-V$ curves are modulated by magnetic field. **(b)** Voltage $V$ at zero bias current, as a function of magnetic flux $\Phi/\Phi_0 = B/\Delta B$ ($\Delta B$ is the oscillation period of the SQUID), plotted for SQUID A, B, C and D. Due to instrument-related factors, the measured $V$ can drift randomly over time during long measurements, as observed in SQUID B, C and D. However, only in SQUID A, we observe the drift in zero-bias $V$ varies periodically with $\Phi$. **(c)** Zero-bias $V$ measured from a resistor. $V$ fluctuates randomly over time between $\pm 10$ μV, which is likely due to the fluctuations in our voltage preamplifiers.

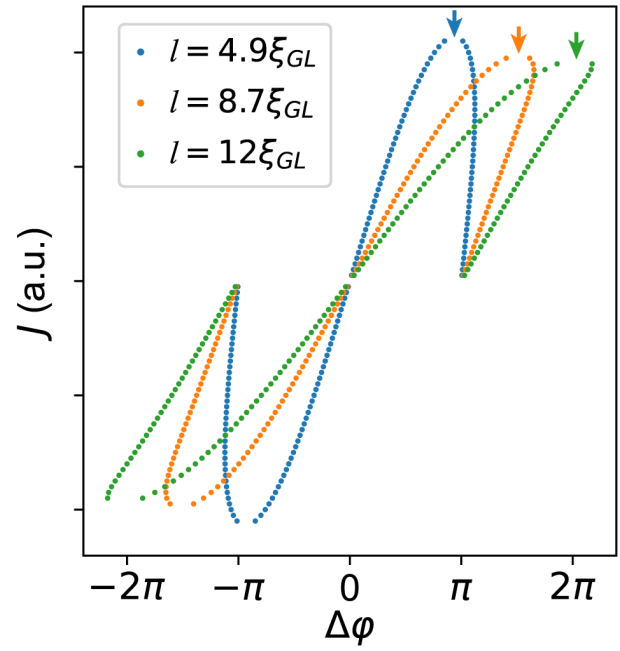

Figure S18. Current-phase relationship of Dayem bridges with various lengths $l$, derived from the theory model. The arrows indicate the points where $\Delta\varphi$ reaches the critical phase $\varphi_c$ and $J$ reaches the critical current density $J_c$.